\documentclass[preprint,aps,prd,showpacs,nofootinbib,superscriptaddress,12pt]{revtex4-1}
\usepackage{amssymb}

\usepackage{graphicx}
\graphicspath{{figures/}{fig/}}
\usepackage{amsmath}
\usepackage{bm}
\usepackage{slashed}
\usepackage{epsfig}
\usepackage{amsfonts}
\usepackage{epstopdf}
\usepackage{color}
\usepackage{extarrows}
\usepackage{multirow}
\usepackage[utf8]{inputenc}
\usepackage{hyperref}

\newcommand{\ben}{\begin{eqnarray}}
\newcommand{\een}{\end{eqnarray}}

\newcommand{\bef}{\begin{figure}[!htp]}
\newcommand{\eef}{\end{figure}}

\newcommand{\bea}{\begin{eqnarray}}
\newcommand{\eea}{\end{eqnarray}}

\def\ba{\begin{linenomath*}\begin{equation}}
\def\ea{\end{equation}\end{linenomath*}}

\allowdisplaybreaks

\begin{document}

\title{\boldmath Evolution of unpolarized transverse momentum
dependent parton distributions of twist-3}


\author{Xue-Tao Liu}
\affiliation{ College of Physics and Communication Electronics, Jiangxi Normal University, Nanchang 330022, China}
\author{Ping-An Liu}
\affiliation{ College of Physics and Communication Electronics, Jiangxi Normal University, Nanchang 330022, China}
\author{Yu-Lu Liu}
\affiliation{ College of Physics and Communication Electronics, Jiangxi Normal University, Nanchang 330022, China}
\author{An-Ping Chen}
\email{chenanping@jxnu.edu.cn}
\affiliation{ College of Physics and Communication Electronics, Jiangxi Normal University, Nanchang 330022, China}
\affiliation{ Jiangxi Provincial Key Laboratory of Advanced Electronic Materials and Devices, Nanchang 330022, China}





\begin{abstract}
 We revisit the calculation of the evolution equations for four unpolarized twist-3 transverse momentum dependent (TMD) parton distribution functions (PDFs) at $\mathcal {O}(\alpha_s)$. Unlike the existing calculations in the literature, which are based on the background field method, we derive the evolution equations directly through diagram expansion. Additionally, instead of using the $\delta$ regulator, we employ the exponential regulator to regularize the rapidity divergences. We find that the evolutions of the four twist-3 TMD PDFs are governed by eight homogeneous equations. Our evolution kernels agree with one set of results in the literature, but differ from another by a sign. Considering the advantages of the exponential regulator in high-order perturbative calculations, our results are crucial for computing the twist-3 TMD PDFs at $\mathcal {O}(\alpha_s^2)$.
\end{abstract}

\maketitle
\allowdisplaybreaks

\section{Introduction}

The Transverse-Momentum-Dependent (TMD) distributions, such as TMD Parton Distribution Functions (TMD PDFs) and TMD Fragmentation Functions (TMD FFs), are non-perturbative and universal quantities that arise from TMD factorization theorems \cite{Ji:2004wu,Ji:2004xq}. These distributions play a critical role in determining the three-dimensional structure of hadrons.
At the twist-2 or leading power (LP) level, TMD factorization has been established for several processes, such as Semi-Inclusive Deeply Inelastic Scattering (SIDIS) \cite{Ji:2004wu,Collins:2004nx}, Drell-Yan processes \cite{Collins:1984kg,Ji:2004xq} and $e^{+}e^{-}$ collisions \cite{Collins:2011zzd}.
The twist-2 TMD distributions have been extensively
studied. It is well-known that twist-2
TMD distributions suffer from both ultraviolet (UV) divergences and rapidity divergences if they are defined with light-cone gauge
links. The UV divergences can be regulated by using
dimensional regularization with $d=4-2\epsilon$ dimensions, which give rise to the renormalization
scale $\mu$. However, rapidity divergences
are not regularized by dimensional regularization, and a dedicated regulator is required, generically denoted as $\tau$. This introduces the rapidity scale $\nu$ dependence in TMD distributions.

In the literature, various rapidity regulators have been proposed to regularize the rapidity divergences.
In Refs. \cite{Collins:1981uw,Ji:2004wu,Ji:2004xq}, the gauge links off light-cone are employed in the definition of TMD distributions. In \cite{Collins:2011zzd} a subtraction method is used to regulate the rapidity divergences, in which the gauge links in the TMD distributions are along light-cone directions, and the rapidity divergences are subtracted by a soft factor, which is defined with gauge links along light-cone directions and those off light-cone. In \cite{Becher:2010tm,Becher:2011dz,Chiu:2011qc,Chiu:2012ir} variations of
the analytic regulator were used. In \cite{Echevarria:2011epo,Echevarria:2012js,Deng:2022gzi}, the $\delta$ regulator was used, in which finite imaginary parts in certain
propagators were used to regulate the rapidity divergences. And in \cite{Li:2016axz} the so-called exponential regulator was proposed. For more details about these different rapidity regulators, please refer to Refs. \cite{Boussarie:2023izj,Ebert:2019okf}. Using an analytic regulator, $\delta$ regulator, and exponential regulator, the quark and gluon TMD PDFs and FFs have been perturbatively calculated at next-to-next-to-leading order (NNLO) \cite{Gehrmann:2012ze,Gehrmann:2014yya,Echevarria:2016scs,Luo:2019hmp,Luo:2019bmw}. Especially, using an exponential regulator, the calculations of quark and gluon TMD PDFs and FFs have been carried out at next-to-next-to-next-to-leading order ($\rm{N^3LO}$) \cite{Luo:2019szz,Luo:2020epw,Ebert:2020yqt,Ebert:2020qef}. On the other hand,
the presence of scales $\mu$ and $\nu$ leads to two evolution equations for twist-2 TMD distributions: the renormalization group (RG) evolution equations and rapidity scale evolution equations\textemdash the Collins-Soper (CS) equations. The evolution kernel for the RG equation has been computed to three-loop accuracy \cite{Gehrmann:2010ue}, while the CS equation kernel has been calculated to four-loop order \cite{Duhr:2022yyp,Moult:2022xzt}.

Compared to twist-2 TMD factorization, research on twist-3 or next-to-leading power (NLP) TMD factorization is much less developed.
In the early works
of \cite{Mulders:1995dh,Boer:2003cm,Bacchetta:2004zf,Bacchetta:2006tn}, a comprehensive tree-level factorization formalism for the SIDIS process, expressed in terms of both twist-2 and twist-3 TMD distributions, was established. This tree-level methodology was further applied to $e^+e^-$ annihilation in \cite{Boer:1997mf} and the Drell-Yan process in \cite{Lu:2011th}. Early studies beyond tree-level were carried out in Refs. \cite{Bacchetta:2008xw,Chen:2016hgw,Bacchetta:2019qkv}. In \cite{Bacchetta:2008xw,Bacchetta:2019qkv} the authors considered the matching between the TMD and collinear descriptions in the intermediate transverse momentum region at the twist-3 level. In \cite{Chen:2016hgw} the authors performed a calculation of the unpolarized twist-3 TMD PDF at NLO ($\mathcal {O}(\alpha_s)$) and derived the Collins-Soper equation for it. In their study the rapidity divergences were regularized by using off-light-cone gauge links. In recent years, a complete expression for twist-3 TMD factorization beyond the tree level has been derived by different groups using various methods. These include the background field method \cite{Vladimirov:2021hdn}, the Soft-Collinear Effective Theory \cite{Ebert:2021jhy}, and the Collins-Soper-Sterman factorization formalisms \cite{Gamberg:2022lju}. Furthermore, using the background field method, the authors  studied the RG evolution equations and the Collins-Soper equations of TMD distributions at twist-3 in Refs. \cite{Vladimirov:2021hdn,Rodini:2022wki,Rodini:2023plb}, where the RG evolution equations are discussed for the first time. In their studies, the $\delta$ regulator is used to regularize the rapidity divergences. However, there is a sign difference between the RG evolution kernels in \cite{Rodini:2022wki} and those in \cite{Rodini:2023plb}. Therefore, an independent calculation of the evolution equations is essential.

In this paper, we revisit the calculation of the RG evolution equations and the Collins-Soper equations for twist-3 TMD PDFs at $\mathcal {O}(\alpha_s)$. We consider the unpolarized twist-3 quark-gluon-quark correlators in this work. Compared to the calculations in Refs. \cite{Vladimirov:2021hdn,Rodini:2022wki,Rodini:2023plb}, we derive the evolution equations straightforwardly by the diagram expansion. On the other hand, in our calculation we employ the exponential regulator to regularize the rapidity divergences. Compared to other regulators, one of the advantages of the exponential regulator is that it is applied to the total momentum of the extra emissions in the final state. Aside from this last integration, the regulator does not alter the structure of the (cut)-propagators in the amplitudes. Because of this, one can employ many modern techniques, such as the integration-by-parts method and differential equations, to perform the loop integration \cite{Luo:2019hmp}. This allows one to derive the twist-2 TMD PDFs and FFs at high orders, as shown in \cite{Luo:2019hmp,Luo:2019bmw,Luo:2019szz,Luo:2020epw,Ebert:2020yqt,Ebert:2020qef}. And this method can also be extended to the NNLO calculation of the twist-3 TMD PDFs. However, before proceeding with such calculations, it is necessary to first derive the RG evolution equations for these distributions.

The rest of the paper is organized as follows. In section \ref{sec:PDF} we give the definition and parametrization of the unpolarized twist-3 quark-gluon-quark correlation function. In section \ref{sec:eom} we use the equation of motion to separate the contributions from quark-quark correlation function at different twists. In section \ref{sec:EQ} we give the details to calculate the RG evolution equations and the Collins-Soper equations for twist-3 TMD PDFs at $\mathcal {O}(\alpha_s)$. We perform the soft subtraction for twist-3 TMD PDFs and derive the evolution equations for the subtracted TMD PDFs in section \ref{sec:sub}, and section \ref{sec:summary} contains our summary. Finally, we list the evolution equations in different support domains in appendix \ref{ap:expression}.

\section{Definition and parametrization}\label{sec:PDF}

In this work we use the light-cone coordinate system. In this system a vector $a^\mu$ is expressed as $a^\mu=(a^+, a^-, a_\perp) =((a^0+a^3)/\sqrt{2}, (a^0-a^3)/\sqrt{2}, a^1, a^2)$ and $a_\perp^2=-(a^1)^2-(a^2)^2$. We introduce two light-like vectors $n^\mu=(0,1,0,0)$ and $l^\mu=(1,0,0,0)$. Using them we further define two transverse tensors $g_{\perp}^{\mu\nu}=g^{\mu\nu}-n^\mu l^\nu-n^\nu l^\mu$ and $\epsilon_{\perp}^{\mu\nu}=\epsilon^{\alpha\beta\mu\nu}l_\alpha n_\beta$. The transverse part of any vector $a^\mu$ can then be obtained as $a_\perp^\mu = g_{\perp}^{\mu\nu} a_\nu$. We further define $a_T^2=-a_\perp^2$.

We consider the unpolarized twist-3 quark-gluon-quark distribution correlation function. The quark-gluon-quark correlation functions were first introduced in \cite{Mulders:1995dh}, and have been discussed in many
works. They are essential for establishing TMD factorization at NLP accuracy. These functions provide a unique theoretical framework for investigating the physics of quark-gluon-quark correlations, offering novel insights into the hadronic partonic structure\textemdash a domain that remains largely unexplored. Such correlations may be considered quantum interference effects, and they could be related to average transverse forces acting on partons inside hadrons \cite{Boussarie:2023izj}. The definition of unpolarized twist-3 quark-gluon-quark distribution is given by \cite{Mulders:1995dh,Boer:2003cm,Bacchetta:2006tn,Gamberg:2022lju}
\begin{align}\label{eq:PhiF}
 \Phi_{F,ij}^\mu (x_1,x_2,k_{3\perp}) =& ig_s \int\frac{d\eta}{2\pi} \frac{d^3\xi}{(2\pi)^3} e^{- i \xi\cdot k_3 + i\eta n\cdot k_2 } \nonumber\\
&\hspace{-2cm} \times
\langle h_A \vert \bar \psi_j( \xi  ) {\mathcal L}_n^\dag (\xi) {\mathcal L}_n ( \eta n)  G^{+ \mu} (\eta n  ) {\mathcal L}_n^\dag ( \eta n){\mathcal L}_n ( 0) \psi_i ( 0 ) \vert h_A \rangle \Big\vert_{\xi^+=0},
\end{align}
where $h_A$ is an unpolarized hadron with momentum $P_A^\mu=(P_A^+,0,0,0)$, $\psi$ is the quark field, $G^{+ \mu}$ is the gluon field-strength tensor, and ${\mathcal L}_n$ is the gauge link along the direction $n^\mu$:
\begin{align}
{\mathcal L}_n ( \xi) =P \textrm{exp}\Big(-ig_s \int_{0}^{\infty} d\lambda  n \cdot A(\lambda n+\xi) \Big).
\end{align}
The sign convention for the strong coupling is $D^\mu=\partial^\mu+ig_s A^\mu$. Here we have neglected the transverse gauge links \cite{Ji:2002aa,Belitsky:2002sm}, as in Feynman gauge whereby the
gluon field strength vanishes at infinity. However, the transverse gauge links are important in certain singular
gauges, like the light-cone gauge \cite{Ji:2002aa,Belitsky:2002sm,Cherednikov:2007tw,Idilbi:2010im}.
In the above definition, $k_3$ is the momentum of the quark that is isolated on one side of the cut, $x_1$ is the momentum fraction associated with the other quark: $k_1^+=x_1P_A^+$, and
$x_2$ denotes the momentum fraction of the gluon: $k_2^+=x_2P_A^+$. These momentum satisfy $k_3=k_2+k_1$, and we denote $k_3^+=x_3 P_A^+$.

The parametrization for the correlation function \eqref{eq:PhiF} is given in Refs. \cite{Bacchetta:2006tn,Ebert:2021jhy,Gamberg:2022lju,Rodini:2022wki}, which reads as
\begin{align}\label{eq:dec}
&\Phi_{F}^{\mu} (x_1,x_2,k_{3\perp}) \nonumber\\
=& \frac{ x_3 M_A}{2} \Big[  (f^\perp-i g^\perp)\frac{k_{3\perp\nu}}{M_A}(g_\perp^{\mu\nu}- i\epsilon_\perp^{\mu\nu}\gamma_5)\frac{\gamma^-}{2}
+ ( f^\perp+i g^\perp)\frac{k_{3\perp\nu}}{M_A}(g_\perp^{\mu\nu}+ i\epsilon_\perp^{\mu\nu}\gamma_5)\frac{\gamma^-}{2}
\nonumber\\
&+ i h\frac{\gamma_\perp^\mu\gamma^-}{2}+ i h^\perp\frac{\gamma_{\perp\nu} \gamma^-}{2}\Big(2g_{\perp}^{\nu\rho}g_\perp^{\mu\sigma}
-g_{\perp}^{\nu\mu}g_\perp^{\rho\sigma} \Big) \frac{k_{3\perp\rho} k_{3\perp\sigma}}{M_A^2}\Big].
\end{align}
The twist-3 TMD PDFs on the rhs depend on $x_1$, $x_2$ and $k_{3T}^2$. In above parametrization $M_A$ is the mass of hadron $h_A$. In general the distribution functions $f^\perp$, $g^\perp$, $h$ and $h^\perp$ are complex. We define the corresponding distributions in $b$-space by following Fourier transforming,
\begin{align}
\tilde{\Phi}_{F}^{\mu} (x_1,x_2,b_{\perp})
=& \int d^2k_{3\perp} e^{i k_{3\perp} \cdot b_\perp} \Phi_{F}^{\mu} (x_1,x_2,k_{3\perp}).
\end{align}
Using \cite{Ebert:2021jhy}
\begin{subequations}
\begin{align}
 &\int d^2k_{\perp} e^{i k_{\perp} \cdot b_\perp} \frac{k_\perp^\mu}{M_A}f(k_{\perp}^2)=-iM_A b_\perp^\mu \tilde{f}^{(1)}(b_\perp^2),\\
 &\int d^2k_{\perp} e^{i k_{\perp} \cdot b_\perp} \frac{k_\perp^\mu k_\perp^\nu}{M_A^2}f(k_{\perp}^2)=\frac{(-i)^2}{2}M_A^2 b_\perp^\mu b_\perp^\nu \tilde{f}^{(2)}(b_\perp^2) +(-i)^2 g_\perp^{\mu\nu}\tilde{f}^{(1)}(b_\perp^2),
\end{align}
\end{subequations}
we then obtain the parametrization in $b$-space as
\begin{align}\label{eq:dec-b}
&\tilde{\Phi}_{F}^{\mu} (x_1,x_2,b_{\perp}) \nonumber\\
=& \frac{ x_3 M_A}{2} \Big[  (\tilde{f}^{\perp(1)}-i\tilde{g}^{\perp(1)})(-iM_A b_{\perp\nu})(g_\perp^{\mu\nu}- i\epsilon_\perp^{\mu\nu}\gamma_5)\frac{\gamma^-}{2}
+ (\tilde{f}^{\perp(1)}+i\tilde{g}^{\perp(1)})(-iM_A b_{\perp\nu})
\nonumber\\
&
\times(g_\perp^{\mu\nu}+ i\epsilon_\perp^{\mu\nu}\gamma_5)\frac{\gamma^-}{2}
+ i\tilde{h}\frac{\gamma_\perp^\mu\gamma^-}{2}+ i\tilde{h}^{\perp(2)}\frac{\gamma_{\perp\nu} \gamma^-}{2}\Big(2g_{\perp}^{\nu\rho}g_\perp^{\mu\sigma}
-g_{\perp}^{\nu\mu}g_\perp^{\rho\sigma} \Big)\nonumber\\
 &\times \frac{(-i)^2}{2}M_A^2b_{\perp\rho} b_{\perp\sigma}\Big].
\end{align}
Following Refs. \cite{Boer:2011xd,Ebert:2021jhy}, here we defined the derivative of the TMD PDFs as
\begin{align}
\tilde{f}^{(n)} (x_1,x_2,b_{\perp})
=& n! \Big(-\frac{2}{M_A^2}\frac{\partial}{\partial b_T^2}\Big)^n \tilde{f} (x_1,x_2,b_{\perp}).
\end{align}

Based on above parametrization, the relevant traces of the quark-gluon-quark correlator are given by
\begin{subequations}\label{eq:tr}
\begin{align}
  &\frac{1}{2x_3} \textrm{Tr}[ \Phi_{F\nu} (x_1,x_2,k_{3\perp}) (g_\perp^{\mu\nu}+ i\epsilon_\perp^{\mu\nu}\gamma_5)\gamma^+] = k_{3\perp}^\mu(f^\perp-
  ig^\perp) ,\\
 & \frac{1}{2x_3} \textrm{Tr}[ \Phi_{F\nu} (x_1,x_2,k_{3\perp}) (g_\perp^{\mu\nu}- i\epsilon_\perp^{\mu\nu}\gamma_5)\gamma^+] = k_{3\perp}^\mu(f^\perp+
  ig^\perp) ,\\
 & \frac{1}{(d-2)x_3M_A} \textrm{Tr}[ \Phi_{F\nu} (x_1,x_2,k_{3\perp}) (i\gamma_{\perp}^\nu\gamma^+ )] = h,\\
 &\frac{M_A}{(d-2)x_3} \textrm{Tr}[\Phi_{F\mu}(x_1,x_2,k_{3\perp}) (i\gamma_{\perp\nu}\gamma^+ )] \Big( (d-2)g_{\perp}^{\nu\rho}g_\perp^{\mu\sigma}
-g_{\perp}^{\nu\mu}g_\perp^{\rho\sigma} \Big) \nonumber\\
&= \Big( (d-2)g_{\perp}^{\nu^{\prime}\rho} g_{\perp}^{\mu^{\prime} \sigma}
-g_{\perp}^{\nu^{\prime}\mu^{\prime}} g_{\perp}^{ \rho \sigma} \Big) k_{3\perp\nu^\prime} k_{3\perp\mu^\prime}h^\perp.
\end{align}
\end{subequations}
Here $d=4-2\epsilon$ is the space-time dimension. These TMD PDFs depend not only on momentum fractions, but also on the renormalization scale $\mu$ and rapidity scale $\nu$ implicitly. The $\mu$ and $\nu$-dependence for the TMD PDFs in Eq.\eqref{eq:tr} is the subject of this work. It should be noted that the TMD PDFs of twist-3 are indefinite at $x_i = 0$, $(i=1,2,3)$ as demonstrated in \cite{Rodini:2022wki}. Nonetheless, they are integrable at $x_i=0$ \cite{Rodini:2022wki}. For simplicity, in the following derivation we assume $x_i>0$. The evolution equations in other support domains can be calculated similarly, and we present them in the appendix. In our following calculation of the evolution equations we will use the Feynman gauge.

\section{Equation of motion relations}\label{sec:eom}

\begin{figure}[htb!]
\begin{center}
\includegraphics[width=0.9\textwidth]{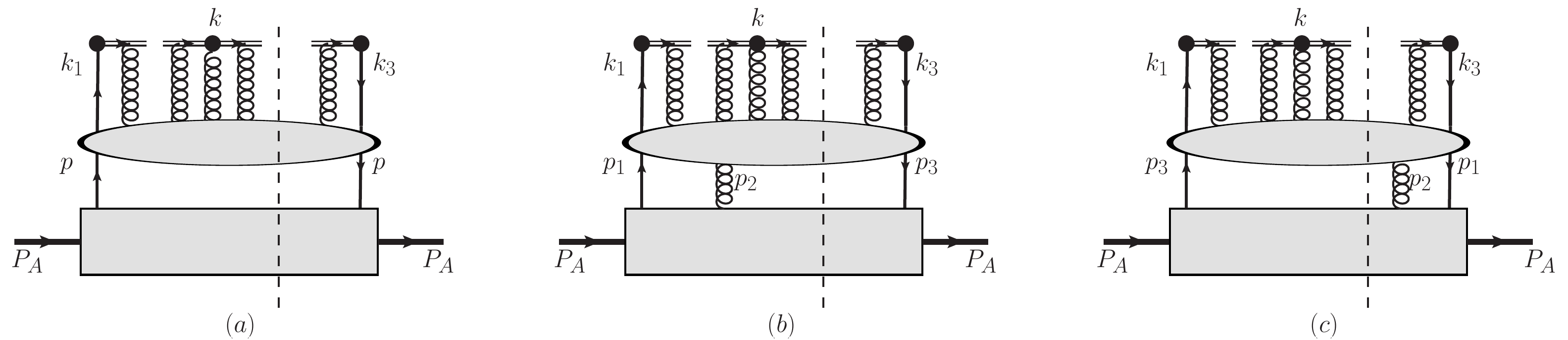}
\end{center}
\caption{The generic structure of diagrams for evolutions.}\label{fig:evo0}
\end{figure}
In order to derive the evolution equations of the twist-3 TMD PDFs, we essentially need to consider the contribution of Fig.\ref{fig:evo0}. Fig.\ref{fig:evo0}(a), Fig.\ref{fig:evo0}(b) and Fig.\ref{fig:evo0}(c) stand for the generic structure of diagrams for the contributions from the quark-quark correlation function and quark-gluon-quark correlation function, respectively.
In the diagrams the black solid circles stand for the insertion of the field operators with given projections of $\gamma$-matrices in the definitions of twist-3 TMD PDFs in Eq.\eqref{eq:tr}. The double line represents the gauge link. The middle bubbles represent various parton scatterings. The boxes in the lower
part of the diagrams in Fig.\ref{fig:evo0}(b) and Fig.\ref{fig:evo0}(c) denote the quark-gluon-quark correlation function and its hermitian conjugate, respectively. And that in Fig.\ref{fig:evo0}(a) denotes the quark-quark correlation function. The quark-quark correlation function is defined as
\begin{align}\label{eq:q-q}
 \Phi_{ij} (x,p_{\perp}) =&  \int \frac{d^3\xi}{(2\pi)^3} e^{- i \xi\cdot p  }
\langle h_A \vert \bar \psi_j( \xi  ) {\mathcal L}_n^\dag (\xi) {\mathcal L}_n ( 0)  \psi_i ( 0 ) \vert h_A \rangle \Big\vert_{\xi^+=0},
\end{align}
here $p$ is the momentum carried by the quark, and $x$ is the corresponding momentum fraction: $p^+=x P_A^+$. It should be noted that the quark-quark correlation function contains the contributions at different twists. Following the discussion in Refs. \cite{Ma:2017upj,Gamberg:2022lju}, to separate the contributions at different twists, we decompose the quark field $\psi$ into the good and bad components:
\begin{align}\label{eq:good}
\psi(\xi)=\psi_+(\xi)+\psi_-(\xi), \quad \psi_+(\xi)=\frac{1}{2}\gamma^-\gamma^+\psi(\xi), \quad \psi_-(\xi)=\frac{1}{2}\gamma^+\gamma^-\psi(\xi),
\end{align}
here $\psi_+ $ is the good component, which is not power suppressed, while $\psi_- $ is the bad component and it is power suppressed compared to $\psi_+ $. Inserting the decomposition Eq.\eqref{eq:good} into Eq.\eqref{eq:q-q}, we separate the contributions at twist-2 and twist-3. The twist-2 quark-quark correlation function, which contains only good components, is given by
\begin{align}\label{eq:q-q-LP}
 \Phi^{\textrm{LP}}_{ij} (x,p_{\perp}) =&  \int \frac{d^3\xi}{(2\pi)^3} e^{- i \xi\cdot p  }
\langle h_A \vert \bar \psi_{+j}( \xi  ) {\mathcal L}_n^\dag (\xi) {\mathcal L}_n ( 0) \psi_{+i} ( 0 ) \vert h_A \rangle \Big\vert_{\xi^+=0}.
\end{align}
Its parametrization is well known \cite{Bacchetta:2006tn,Ebert:2021jhy,Gamberg:2022lju,Rodini:2022wki}:
\begin{subequations}\label{eq:qq-dec}
\begin{align}
   \frac{1}{2}\textrm{Tr}[ \Phi^{\textrm{LP}} (x,p_{\perp}) \gamma^+] =&  f_1(x,p_{\perp}),\\
   \frac{1}{2}\textrm{Tr}[ \Phi^{\textrm{LP}} (x,p_{\perp})  (i\sigma^{\alpha+} \gamma^5)] =&  -\frac{\epsilon_\perp^{\alpha\rho}p_{\perp\rho}}{M_A}h_1^\perp(x,p_{\perp}).
\end{align}
\end{subequations}
The correlation functions which contain one good field and one bad field are power suppressed compared to $\Phi^{\textrm{LP}}$. We refer to their sum as a twist-3 quark-quark correlation function, which is given by
\begin{align}\label{eq:q-q-NLP}
\Phi^{\textrm{NLP}} (x,p_{\perp})=& \Phi^{\textrm{NLP},A} (x,p_{\perp})+\Phi^{\textrm{NLP},B} (x,p_{\perp}),\nonumber\\
 \Phi^{\textrm{NLP},A}_{ij} (x,p_{\perp}) =&  \int \frac{d^3\xi}{(2\pi)^3} e^{- i \xi\cdot p  }
\langle h_A \vert \bar \psi_{+j}( \xi  ) {\mathcal L}_n^\dag (\xi) {\mathcal L}_n ( 0) \psi_{-i} ( 0 ) \vert h_A \rangle \Big\vert_{\xi^+=0}, \nonumber\\
\Phi^{\textrm{NLP},B}_{ij} (x,p_{\perp}) =&  \int \frac{d^3\xi}{(2\pi)^3} e^{- i \xi\cdot p  }
\langle h_A \vert \bar \psi_{-j}( \xi  ){\mathcal L}_n^\dag (\xi) {\mathcal L}_n ( 0)  \psi_{+i} ( 0 ) \vert h_A \rangle \Big\vert_{\xi^+=0}.
\end{align}
Finally, the correlation function that contains only bad components is twist-4 correlation function, it is irrelevant for our study.

Using the equation of motion: $2D^+\psi_-+\gamma^+\gamma_\perp\cdot D_\perp \psi_+=0$ \cite{Bacchetta:2006tn,Gamberg:2022lju}, one can express the bad field in the correlation functions in Eq.\eqref{eq:q-q-NLP} with the good field combined with gluon fields. Take $\Phi^{\textrm{NLP},A}$ as an example, we have \cite{Ma:2017upj,Gamberg:2022lju}
\begin{align}\label{eq:q-q-NLP-A}
 & \Phi^{\textrm{NLP},A}_{ij} (x,p_{\perp}) \nonumber\\
=&\int \frac{d^3\xi}{(2\pi)^3} e^{- i \xi\cdot p  }
\langle h_A \vert \bar \psi_{+j}( \xi  ) {\mathcal L}_n^\dag (\xi) {\mathcal L}_n ( 0)  \Big(\frac{i\gamma^+}{2p^+} (i\gamma_\perp\cdot p_\perp) \psi_+ ( 0 )\Big)_i \vert h_A \rangle \Big\vert_{\xi^+=0}
 \nonumber\\
& + \int dp_2^+\frac{i}{p_2^+}
 \int\frac{d\eta}{2\pi} \frac{d^3\xi}{(2\pi)^3} e^{- i \xi\cdot p + i\eta n\cdot p_2 }
 \nonumber\\
&\times
\langle h_A \vert \bar \psi_{+j}( \xi  )  {\mathcal L}_n^\dag (\xi) {\mathcal L}_n ( \eta n) G^{a,+\mu} (\eta n  ) {\mathcal L}_n^\dag ( \eta n){\mathcal L}_n ( 0) \Big(\frac{i\gamma^+}{2p^+} (-ig_s\gamma_\mu T^a) \psi_+ ( 0 )\Big)_i \vert h_A \rangle \Big\vert_{\xi^+=0}.
\end{align}
We find the first term on the rhs is the twist-2 quark-quark correlation function $\Phi^{\textrm{LP}}$, but contains a power correction associated with $p_\perp$. And
the second term is the twist-3 quark-gluon-quark correlation function.

\begin{figure}[htb!]
\begin{center}
 \includegraphics[width=0.9\textwidth]{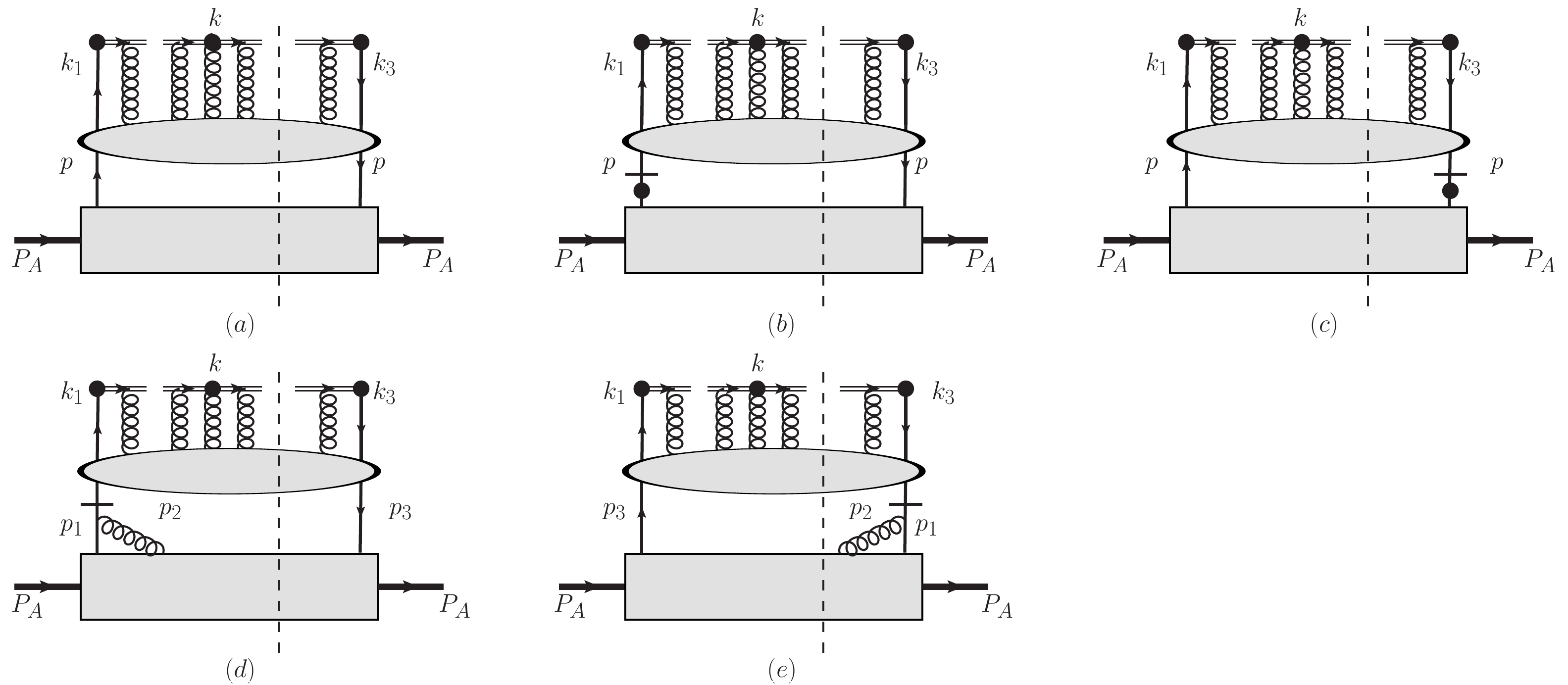}
 \end{center}
 \caption{Patterns of diagrams representing contributions from the quark-quark correlation function after applying the equation of motion.}\label{fig:evo1}
\end{figure}

Using the relations in Eqs. \eqref{eq:q-q-LP}, \eqref{eq:q-q-NLP} and \eqref{eq:q-q-NLP-A}, we are now able to correctly separate contributions at different twists from Fig.\ref{fig:evo0}(a). The contribution from twist-2 correlation functions can be obtained by calculating Fig.\ref{fig:evo0}(a) with the bottom-box replaced by
the $ \Phi^{\textrm{LP}}$, as shown in Fig.\ref{fig:evo1}(a), where the bottom-box represents $ \Phi^{\textrm{LP}}$. The contribution from twist-3 correlation functions can be obtained by replacing the bottom-box in Fig.\ref{fig:evo0}(a) with the terms on the r.h.s. of Eq. \eqref{eq:q-q-NLP-A}, as showed in Fig.\ref{fig:evo1}(b-e). Fig.\ref{fig:evo1}(b) stands for the contribution from the term in the second line of Eq. \eqref{eq:q-q-NLP-A}, where the bottom-box represents $ \Phi^{\textrm{LP}}$, the quark line with the short bar represents the special propagator $i\gamma^+/2p^+$, and the black dot near the bottom-box denotes the vertex $i\gamma_\perp\cdot p_\perp$. Fig.\ref{fig:evo1}(c) is the contribution from the hermitian conjugate. Fig.\ref{fig:evo1}(d) and (e) stand for the contributions from the term in the last line of Eq. \eqref{eq:q-q-NLP-A}, where the bottom-boxes represent the twist-3 quark-gluon-quark correlation function and its hermitian conjugate, respectively.

\section{Evolution equations}\label{sec:EQ}

With the above discussion, we are now able to calculate the evolution equations of the twist-3 TMD PDFs based on the diagrams in Fig.\ref{fig:evo0}(b), Fig.\ref{fig:evo0}(c) and Fig.\ref{fig:evo1}(a-e). In the calculation, we employ the $\overline{\textrm{MS}}$ scheme to subtract the UV divergences. Consequently, there is a factor $[\mu^2 \textrm{exp}(\gamma_E)/(4\pi)]^\epsilon$ associated with each factor of the strong coupling $\alpha_s = g_s^2/(4\pi)$.

\begin{figure}[htb!]
\begin{center}
 \includegraphics[width=0.9\textwidth]{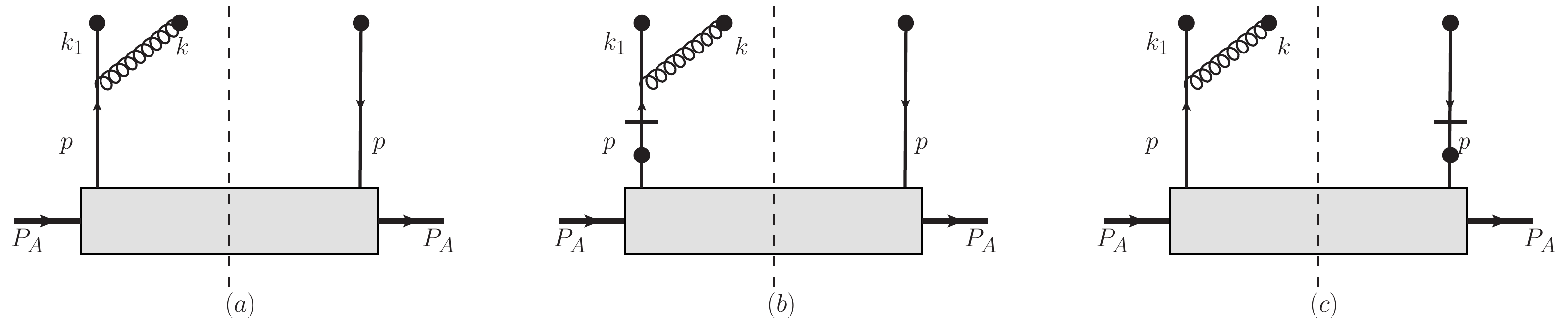}
 \end{center}
 \caption{Diagrams for the contributions from twist-2 quark-quark correlation function.}\label{fig:evolo}
\end{figure}

We take the distribution function $h$ as an example to show the calculation.
The contributions from twist-2 quark-quark correlation function at LO ($\mathcal {O}(\alpha_s)$) are shown in Fig.\ref{fig:evolo}.
Here the real diagrams have no contribution because they have no UV or rapidity divergences. From Fig.\ref{fig:evolo}(a) we have
\begin{align}\label{eq:3a}
h(x_1,x_2,k_{3\perp})\Big\vert_{3a} =& \frac{ig_s}{(d-2) x_3 M_A}\Big(\frac{\mu^2e^{\gamma_E}}{4\pi}\Big)^\epsilon \int dp^+ d^{d-2}p_{\perp} \int\frac{d^{d}k}{(2\pi)^d}\delta(k^+-x_2P_A^+)
\nonumber\\
&\hspace{-2cm}\times \delta(p^+ - x_3 P_A^+)
 \delta^{(d-2)}(p_{\perp}-k_{3\perp})
 \textrm{Tr}\Big[ (i\gamma_\perp^\nu \gamma^+)(-i)(k^+ g_{\nu \mu}-k_\nu n_\mu)T^a
\frac{i( p-k)\cdot \gamma}{( p-k)^2+i\varepsilon}
\nonumber\\
&\hspace{-2cm} \times(-ig_s\gamma_\alpha T^a)
  \frac{-ig^{\mu\alpha}}{k^2+i\varepsilon}
 \mathcal {M}(p) \Big].
\end{align}
Here
\begin{align}
\mathcal {M}_{ij}(p)&=\int \frac{d^{d-1}\xi}{(2\pi)^{d-1}} e^{- i \xi\cdot p  }
\langle h_A \vert \bar \psi_{+j}( \xi  )     \psi_{+i} ( 0 ) \vert h_A \rangle \Big\vert_{\xi^+=0}.
\end{align}
Using the decomposition in Eq.\eqref{eq:qq-dec}, we derive
\begin{align}
h(x_1,x_2,k_{3\perp})\Big\vert_{3a} =& \frac{ig_s^2C_F}{2(d-2) x_3 M_A^2}   \Big(\frac{\mu^2e^{\gamma_E}}{4\pi}\Big)^\epsilon
\int\frac{d^{d}k}{(2\pi)^d}\delta(k^+-x_2P_A^+) \frac{1 }{k^2+i\varepsilon}
 \nonumber\\
&\times \textrm{Tr}\Big[\gamma_\perp^\nu \gamma^+ (k^+ g_{\nu \mu}-k_\nu n_\mu)
\frac{( k_3-k)\cdot \gamma}{( k_3-k)^2+i\varepsilon}
\gamma_\alpha
\gamma_\rho \gamma^- k_{3\perp}^\rho g^{\mu\alpha}\Big]
\nonumber\\
&\times h_1^\perp(x_3,k_{3\perp}) ,
\end{align}
with $k_3^\mu=(x_3P_A^+,0,k_{3\perp})$, $C_F=(N_c^2-1)/(2N_c)$, $C_A=N_c$. We find $f_1$ has no contribution. Expanding the expression in the trace in $k_{3T}^2$ and taking the NLP term, we have
\begin{align}
h(x_1,x_2,k_{3\perp})\Big\vert_{3a} =& \frac{ig_s^2C_F}{2(d-2) x_3 M_A^2}
\Big(\frac{\mu^2e^{\gamma_E}}{4\pi}\Big)^\epsilon   \int\frac{d^{d}k}{(2\pi)^d}
 \delta(k^+-x_2P_A^+)\frac{1 }{k^2+i\varepsilon} h_1^\perp(x_3,k_{3\perp})k_{3T}^2
  \nonumber\\
&\times \Big( \frac{\partial}{\partial k_{3T}^2}\textrm{Tr}\Big[\gamma_\perp^\nu\gamma^+ (k^+ g_{\nu \mu}-k_\nu n_\mu)
\frac{( k_3-k)\cdot \gamma}{( k_3-k)^2+i\varepsilon}
\gamma_\alpha
\gamma_\rho \gamma^- k_{3\perp}^\rho g^{\mu\alpha} \Big]
\Big\vert_{k_{3\perp}=0}\Big)\nonumber\\
=& \frac{ig_s^2C_F}{x_3 M_A^2}\Big(\frac{\mu^2e^{\gamma_E}}{4\pi}\Big)^\epsilon h_1^\perp(x_3,k_{3\perp})k_{3T}^2 \int\frac{\mathrm{d}^{d}k}{(2\pi)^d}
 \delta(k^+-x_2P_A^+) \nonumber\\
 &\times \frac{2}{d-2}\Big[- \frac{2((d-2)k^+-2k_3^+)k_T^2}{(d-2)(k^2+i\varepsilon)(k^2-2k^-k_3^++i\varepsilon)^2}
 \nonumber\\&-
 \frac{(d-4)k^+}{(k^2+i\varepsilon)(k^2-2k^-k_3^++i\varepsilon)}\Big].
\end{align}
We first carry out the
integration over $k^-$. The denominators contain two singularities in the $k^-$ plane:
\begin{align}
 k^-_{(1)}=\frac{k_T^2-i\varepsilon}{-2(k_3^+  -k^+)}, \quad\quad k^-_{(2)}=\frac{k_T^2-i\varepsilon}{2  k^+}.
\end{align}
Only when $0<k^+<k_3^+$, the singularities are on opposite sides of the real axis, which can give a none vanished value for the contour integral of $k^-$. In this case, we close the integration contour at $\infty$ in the lower half-plane and take the residue at the point $k^- = k^-_{(2)}$. Then we obtain
\begin{align}
h(x_1,x_2,k_{3\perp})\Big\vert_{3a}
=& \frac{2\alpha_sC_F}{x_3 M_A^2} \frac{x_1x_2}{x_3^2}h_1^\perp(x_3,k_{3\perp})k_{3T}^2 \frac{1}{4\pi}\Big(\frac{1}{\epsilon_{\textrm{UV}}} -\frac{1}{\epsilon_{\textrm{IR}}} \Big).
\end{align}
Here we have used
\begin{align}\label{eq:kT}
\mu^{2\epsilon}\int\frac{d^{d-2}k_\perp}{(2\pi)^{d-2}}\frac{1}{k_T^2}=\frac{1}{4\pi}
\Big(\frac{1}{\epsilon_{\textrm{UV}}} -\frac{1}{\epsilon_{\textrm{IR}}} \Big).
\end{align}

In a similar way, we can calculate the contributions from Fig.\ref{fig:evolo}(b) and Fig.\ref{fig:evolo}(c), we obtain
\begin{subequations}
\begin{align}
h(x_1,x_2,k_{3\perp})\Big\vert_{3b}
=&- \frac{2\alpha_sC_F}{x_3 M_A^2} \frac{x_1x_2}{x_3^2}h_1^\perp(x_3,k_{3\perp})k_{3T}^2 \frac{1}{4\pi}\Big(\frac{1}{\epsilon_{\textrm{UV}}} -\frac{1}{\epsilon_{\textrm{IR}}} \Big),\\
h(x_1,x_2,k_{3\perp})\Big\vert_{3c}
=&0.
\end{align}
\end{subequations}
We find that the sum of the three diagrams is zero. Therefore, the contribution from the twist-2 quark-quark correlation function at one-loop vanishes. Consequently, there is no mixing with the quark-quark correlation function at $\mathcal{O}(\alpha_s)$.

\begin{figure}[htb!]
\begin{center}
 \includegraphics[width=0.9\textwidth]{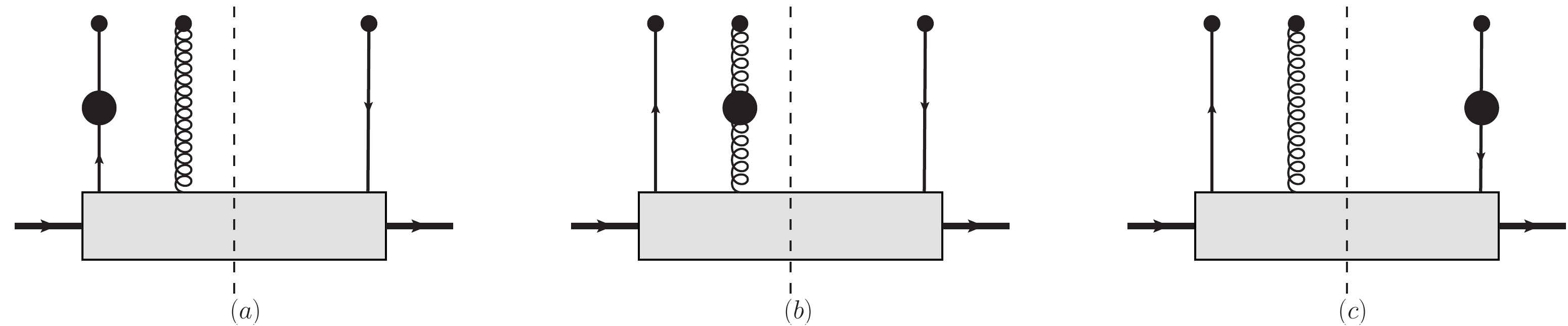}
 \end{center}
 \caption{A set of diagrams for the contributions from twist-3 quark-gluon-quark correlation function. This set only contains the self-energy corrections. The black disk includes the one loop quark or gluon self-energy.}\label{fig:self-energy}
\end{figure}

Now we turn to the contributions from the twist-3 quark-gluon-quark correlation function. The contributions from the quark or gluon self-energy correction are shown in Fig. \ref{fig:self-energy}.
The renormalization factors of these diagrams are well-known. Taking the renormalization of $g_s$ in the definition of $\Phi_F^\mu$ into account, we have
\begin{align}
h(x_1,x_2,k_{3\perp})\Big\vert_{\textrm{Fig.4, UV}}
=&  \frac{\alpha_s}{4\pi}(- C_F -C_A)
\Big(\frac{1}{\epsilon_{\textrm{UV}}} +\ln\mu^2 \Big)
 h(x_1,x_2,k_{3\perp}).
\end{align}
It is convenient to derive the evolution equations in $b$-space. In $b$-space we have
\begin{align}\label{eq:SE}
\tilde{h}(x_1,x_2,b_{\perp})\Big\vert_{\textrm{Fig.4, UV}}
=&  \frac{\alpha_s}{4\pi}(- C_F -C_A)
\Big(\frac{1}{\epsilon_{\textrm{UV}}} +\ln\mu^2 \Big)
 \tilde{h}(x_1,x_2,b_{\perp}).
\end{align}

\begin{figure}[htb!]
\begin{center}
\includegraphics[width=0.9\textwidth]{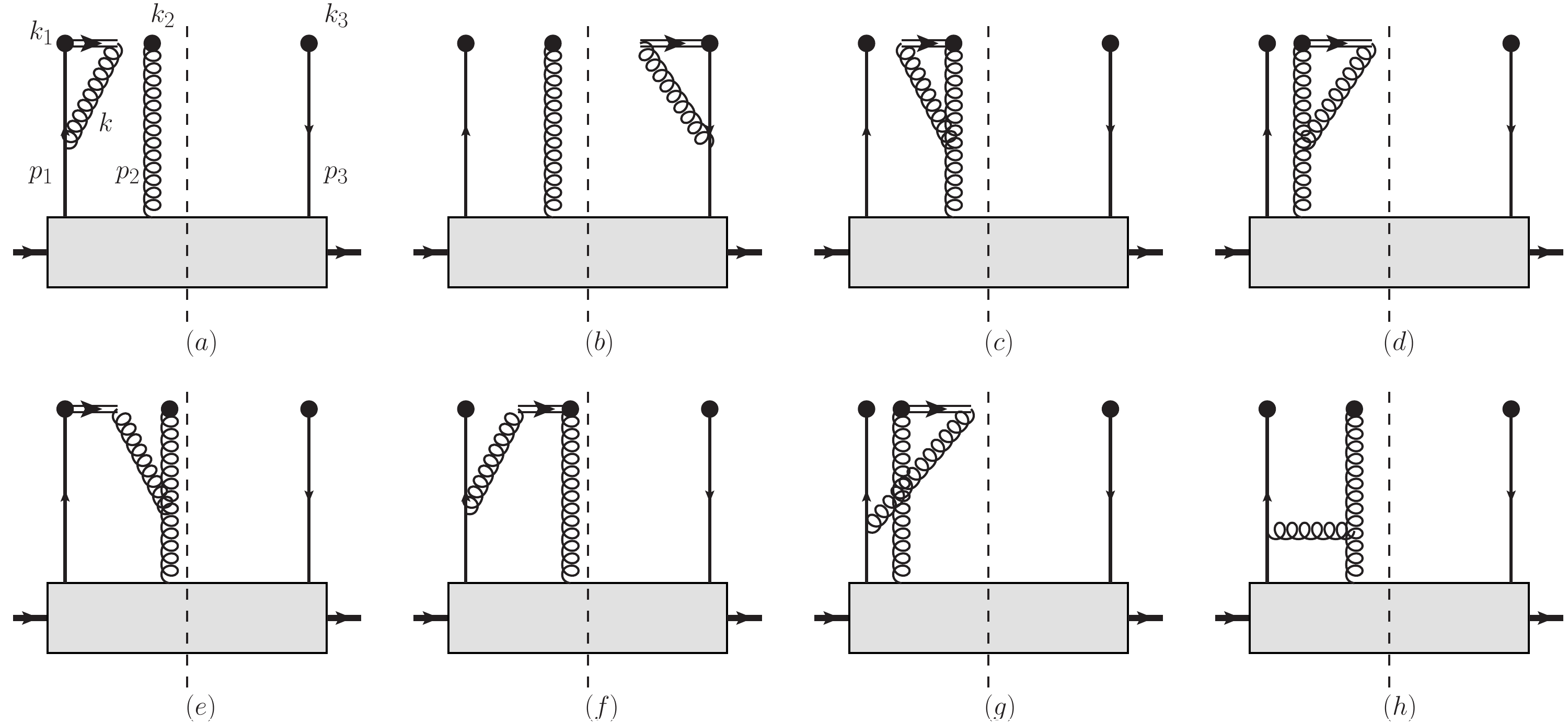}
\includegraphics[width=0.9\textwidth]{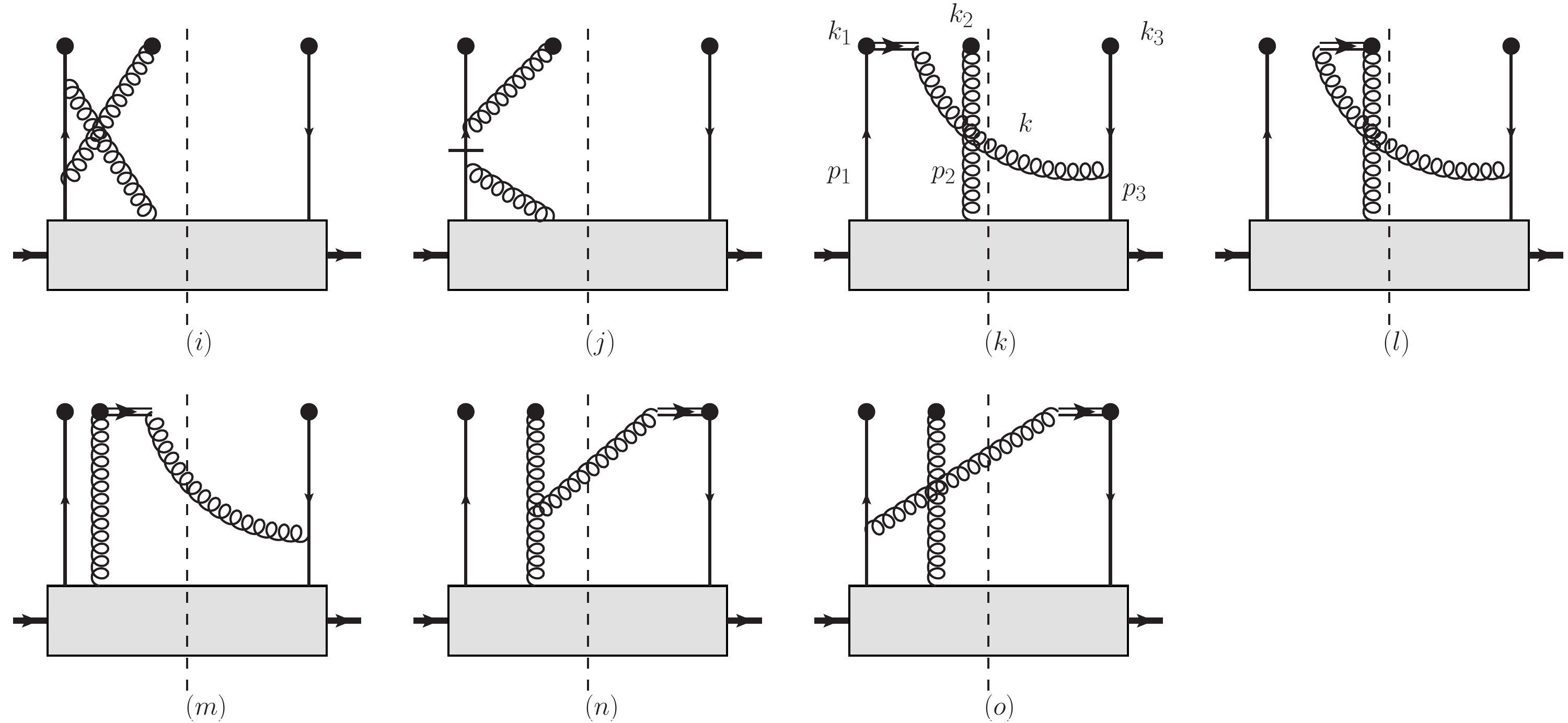}
\end{center}
\caption{Another set of diagrams for the contributions from twist-3 quark-gluon-quark correlation function.}\label{fig:evo}
\end{figure}

The remaining diagrams are given in Fig. \ref{fig:evo}. Here only the diagrams containing UV or rapidity divergences are included. We will first derive the $1/\epsilon_{\textrm{IR}}$ (IR) divergences of these diagrams, where the rapidity divergences associated with the IR divergences do not need to be regularized, as they will cancel out between the virtual and real corrections. Next, we derive the total $1/\epsilon$ divergences and rapidity divergences of these diagrams, where the rapidity divergences are regularized using the exponential regulator. By using the results containing $1/\epsilon$ and rapidity divergences to subtract the results containing $1/\epsilon_{\textrm{IR}}$ divergences, we obtain the $1/\epsilon_{\textrm{UV}}$ (UV) divergences and the regularized rapidity divergences of these diagrams.

We first consider Fig. \ref{fig:evo}(a), we have
\begin{align}\label{eq:AM}
h(x_1,x_2,k_{3\perp})\Big\vert_{5a} =& \frac{ig_s}{(d-2) x_3 M_A} \Big(\frac{\mu^2e^{\gamma_E}}{4\pi}\Big)^\epsilon
\int dp_2^+ d^{d-2}p_{2\perp}  dp_3^+ d^{d-2}p_{3\perp} \int\frac{d^{d}k}{(2\pi)^d}
 \nonumber\\
&\hspace{-2cm} \times \delta(p_2^+-x_2P_A^+)\delta(p_3^+ - x_3 P_A^+) \delta^{(d-2)}(p_{3\perp}-k_{3\perp})
\textrm{Tr}  \Big[ (i\gamma_\perp^\nu \gamma^+)
(-i)(k_2^+g_{\nu\mu}-k_{2\nu} n_\mu)\nonumber\\
&\hspace{-2cm}\times
T^b (-i g_s T^a )
\frac{i n^\rho}{-k^+ +i\varepsilon}  \frac{i( p_1-k)\cdot \gamma}{( p_1-k)^2+i\varepsilon} (-ig_s\gamma_\rho T^a)  \frac{-i}{k^2+i\varepsilon}
 \mathcal {M}_{A}^{b,\mu}(p_2,p_3) \Big].
\end{align}
Here
\begin{align}
\mathcal {M}_{A,ij}^{b,\mu}(p_2,p_3)= &\int\frac{d^{d-1}\zeta}{(2\pi)^{d-1}} \frac{d^{d-1}\xi}{(2\pi)^{d-1}} e^{- i \xi\cdot p_3 + i\zeta \cdot p_2 }
\langle h_A \vert \bar \psi_{+j}( \xi  )    A^{b,\mu} (\zeta  )   \psi_{+i} ( 0 ) \vert h_A \rangle \Big\vert_{\xi^+=0,\zeta^+=0}.
\end{align}
In the above equations, the momentum scales as $p_i^\mu\sim P_A^+(1,\lambda^2,\lambda)$, $k_i^\mu\sim P_A^+(1,\lambda^2,\lambda)$, $(i=1,2,3)$, with $\lambda\sim k_{3\perp}/P_A^+$. The collinear gluon scales as $A^{b,\mu}\sim P_A^+(1,\lambda^2,\lambda)$. At NLP, both $A^{b,+}$ and $A^{b,\mu}_\perp$ have contributions. And they were converted to field strength $G^{b,+\mu}$ through partial integration. To simplify, here we only calculate the contribution of $A^{b}_\perp$. And this should not affect the final result. In calculating the $A^{b}_\perp$ contribution, we expand the expression in the square brackets in $\lambda$ to NLP. We then have
\begin{align}
h(x_1,x_2,k_{3\perp})\Big\vert_{5a} =& \frac{iC_F}{(d-2) x_3 M_A} \Big(\frac{\mu^2e^{\gamma_E}}{4\pi}\Big)^\epsilon
  \int\frac{d^{d}k}{(2\pi)^d}
\textrm{Tr}  \Big[ (i\gamma_\perp^\nu \gamma^+)
(-i)(k_2^+g_{\nu\mu}-\hat k_{2\nu} n_\mu)
\nonumber\\
&\hspace{-2cm}\times (-i g_s  )
\frac{i n^\rho}{-k^+ +i\varepsilon}  \frac{i( \hat p_1-k)\cdot \gamma}{( \hat p_1-k)^2+i\varepsilon} (-ig_s\gamma_\rho )  \frac{-i}{k^2+i\varepsilon}
 \frac{1}{k_2^+} \Phi_F^\mu (x_1,x_2,k_{3\perp}) \Big].
\end{align}
with $\hat{p}_i^\mu=(p_i^+,0,0_\perp)$, $\hat{k}_i^\mu=(k_i^+,0,0_\perp)$, $i=1,2,3$. Inserting the parametrization of $\Phi_F^\mu$ into above equation, we find only $h$ has contribution, which reads as
\begin{align}
h(x_1,x_2,k_{3\perp})\Big\vert_{5a} =& \frac{iC_F}{4(d-2) } \Big(\frac{\mu^2e^{\gamma_E}}{4\pi}\Big)^\epsilon
  \int\frac{d^{d}k}{(2\pi)^d}\frac{1}{k_2^+}
\textrm{Tr}  \Big[ (i\gamma_\perp^\nu \gamma^+)
(-i)(k_2^+g_{\nu\mu}-\hat k_{2\nu} n_\mu)
\nonumber\\
&\hspace{-2cm}\times (-i g_s  )
\frac{i n^\rho}{-k^+ +i\varepsilon}  \frac{i( \hat p_1-k)\cdot \gamma}{( \hat p_1-k)^2+i\varepsilon} (-ig_s\gamma_\rho )  \frac{-i}{k^2+i\varepsilon}
  (i\gamma_\perp^\mu\gamma^-) \Big]h (x_1,x_2,k_{3\perp}).
\end{align}
We first integrate over $k^-$ using contour integration. We then perform the integration over $k_\perp$ using Eq. \eqref{eq:kT}. Finally, we obtain
\begin{align}
h(x_1,x_2,k_{3\perp})\Big\vert_{5a}   =&
 2\alpha_sC_F \Big(1-\ln x_1-\int_0^1dy\frac{1}{y-i\varepsilon} \Big)
 \frac{1}{4\pi}\Big(\frac{1}{\epsilon_{\textrm{UV}}} -\frac{1}{\epsilon_{\textrm{IR}}}\Big)
h(x_1,x_2,k_{3\perp}).
\end{align}
Employing the identity
\begin{align}
\int_0^1dy\frac{1}{y\pm i\varepsilon} =\int_0^1dy\Big(\frac{1}{y}\mp \frac{i\pi}{2}\delta(y)\Big),
\end{align}
we have
\begin{align}
h(x_1,x_2,k_{3\perp})\Big\vert_{5a}   =&
 2\alpha_sC_F \Big(1-\ln x_1-\int_0^1dy\frac{1}{y}-\frac{i\pi}{2} \Big)
 \frac{1}{4\pi}\Big(\frac{1}{\epsilon_{\textrm{UV}}} -\frac{1}{\epsilon_{\textrm{IR}}}\Big)
h(x_1,x_2,k_{3\perp}).
\end{align}
The above result contains a divergence in the integral of $y$. This divergence is the rapidity divergence. In the subsequent derivation, this divergence will be either canceled or properly regulated through the introduction of a rapidity regulator. In $b$-space, we then have
\begin{align}\label{eq:5a}
\tilde{h}(x_1,x_2,b_{\perp})\Big\vert_{5a}   =&
 2\alpha_sC_F \Big(1-\ln x_1-\int_0^1dy\frac{1}{y}-\frac{i\pi}{2} \Big)
 \frac{1}{4\pi}\Big(\frac{1}{\epsilon_{\textrm{UV}}} -\frac{1}{\epsilon_{\textrm{IR}}}\Big)
\tilde{h}(x_1,x_2,b_{\perp}).
\end{align}

Similarly, for Fig. \ref{fig:evo}(b)-(j), we derive
\begin{subequations}\label{eq:fig5v}
\begin{align}
\tilde{h}(x_1,x_2,b_{\perp})\Big\vert_{5b}   =&
 2\alpha_sC_F \Big(1-\ln(x_1+x_2)-\int_0^1dy\frac{1}{y}+\frac{i\pi}{2} \Big)
 \frac{1}{4\pi}\Big(\frac{1}{\epsilon_{\textrm{UV}}} -\frac{1}{\epsilon_{\textrm{IR}}}\Big)
\tilde{h}(x_1,x_2,b_{\perp}),\\
\tilde{h}(x_1,x_2,b_{\perp})\Big\vert_{5c}   =&
 \alpha_s\frac{C_A}{2} \Big(1-2\ln x_2-\int_0^1dy\frac{2}{y}-i\pi \Big)
 \frac{1}{4\pi}\Big(\frac{1}{\epsilon_{\textrm{UV}}} -\frac{1}{\epsilon_{\textrm{IR}}}\Big)
\tilde{h}(x_1,x_2,b_{\perp}),\\
\tilde{h}(x_1,x_2,b_{\perp})\Big\vert_{5d}   =&
 \alpha_s\frac{C_A}{2} \Big(1-2\ln x_2-\int_0^1dy\frac{2}{y}-i\pi \Big)
 \frac{1}{4\pi}\Big(\frac{1}{\epsilon_{\textrm{UV}}} -\frac{1}{\epsilon_{\textrm{IR}}}\Big)
\tilde{h}(x_1,x_2,b_{\perp}),\\
\tilde{h}(x_1,x_2,b_{\perp})\Big\vert_{5e}
 =&
 \alpha_s \frac{C_A}{2} \Big[\int_{-\infty}^\infty dy_2  \theta(y_2-x_2)\frac{x_2(x_2+y_2)}{(y_2-x_2) y_2^2}\tilde{h}(y_1,y_2,b_{\perp}) +i\pi \tilde{h}(x_1,x_2,b_{\perp})\Big]\nonumber\\&\hspace{-2cm}\times
 \frac{1}{4\pi}\Big(\frac{1}{\epsilon_{\textrm{UV}}} -\frac{1}{\epsilon_{\textrm{IR}}} \Big),\\
\tilde{h}(x_1,x_2,b_{\perp})\Big\vert_{5f}
 =&
 2\alpha_s C_F \Big[ \int_{-\infty}^\infty dy_2  \theta(x_2-y_2)\frac{x_1x_2}{(x_2-y_2) y_1y_2}\tilde{h}(y_1,y_2,b_{\perp}) +\frac{i\pi}{2} \tilde{h}(x_1,x_2,b_{\perp}) \Big]
  \nonumber\\
&\hspace{-2cm}\times
 \frac{1}{4\pi}\Big(\frac{1}{\epsilon_{\textrm{UV}}} -\frac{1}{\epsilon_{\textrm{IR}}} \Big),\\
\tilde{h}(x_1,x_2,b_{\perp})\Big\vert_{5g}
 =&
 2\alpha_s \Big(C_F-\frac{C_A}{2}\Big) \Big[ \int_{-\infty}^\infty dy_2  \theta(x_2-y_2)\frac{-x_1x_2}{(x_2-y_2) y_1y_2}  \tilde{h}(y_1,y_2,b_{\perp})
 \nonumber\\
&\hspace{-2cm}-\frac{i\pi}{2} \tilde{h}(x_1,x_2,b_{\perp}) \Big]
 \frac{1}{4\pi}\Big(\frac{1}{\epsilon_{\textrm{UV}}} -\frac{1}{\epsilon_{\textrm{IR}}} \Big)
,\\
 \tilde{h}(x_1,x_2,b_{\perp})\Big\vert_{5h}
 =&
 2\alpha_s \Big(\frac{C_A}{2}\Big) \int_{-\infty}^\infty dy_2 \Big[ \theta(x_2-y_2)\frac{x_1( x_2-y_1)}{(y_1+y_2) y_1y_2}
+ \theta(y_2-x_2)
\nonumber\\
&\hspace{-2cm}\times\frac{x_2( 2x_2-y_1+y_2)}{2(y_1+y_2) y_2^2}  \Big]
 \frac{1}{4\pi}\Big(\frac{1}{\epsilon_{\textrm{UV}}} -\frac{1}{\epsilon_{\textrm{IR}}} \Big)
 \tilde{h}(y_1,y_2,b_{\perp}),\\
 \tilde{h}(x_1,x_2,b_{\perp})\Big\vert_{5i}   =&
 2\alpha_s\Big(C_F - \frac{C_A}{2} \Big) \int_{-\infty}^\infty dy_2 \Big[ \theta(y_2-x_1)\frac{-x_1( x_2-y_1)}{(y_1+y_2) y_2^2}
- \theta(x_1-y_2)
\nonumber\\
&\hspace{-2cm} \times \frac{x_2( x_2-y_1)}{(y_1+y_2) y_1y_2}  \Big]
 \frac{1}{4\pi}\Big(\frac{1}{\epsilon_{\textrm{UV}}} -\frac{1}{\epsilon_{\textrm{IR}}}\Big)
\tilde{h}(y_1,y_2,b_{\perp}),\\
\tilde{h}(x_1,x_2,b_{\perp})\Big\vert_{5j}
 =&
 2\alpha_s C_F \int_{-\infty}^\infty dy_2 \frac{-x_1x_2}{(y_1+y_2)^2y_2}
\frac{1}{4\pi}\Big(\frac{1}{\epsilon_{\textrm{UV}}} -\frac{1}{\epsilon_{\textrm{IR}}} \Big)
 \tilde{h}(y_1,y_2,b_{\perp}).
\end{align}
\end{subequations}
Here we defined $p_1^+= y_1P^+_A$, $p_2^+= y_2P^+_A$. For these virtual diagrams, we have $y_1 + y_2 =x_1 +x_2 = x_3$. The contributions of Fig. \ref{fig:evo}(e)-(g) contain the rapidity divergence at $x_2=y_2$.

The real diagrams Fig. \ref{fig:evo}(k)-(o) have no UV divergence, but they contain rapidity divergences. For Fig. \ref{fig:evo}(k), we have
\begin{align}\label{eq:AMk}
h(x_1,x_2,k_{3\perp})\Big\vert_{5k} =& \frac{ig_s}{(d-2) x_3 M_A}\Big(\frac{\mu^2e^{\gamma_E}}{4\pi}\Big)^\epsilon
 \int dp_2^+ d^{d-2}p_{2\perp}  dp_3^+ d^{d-2}p_{3\perp} \int\frac{d^{d}k}{(2\pi)^d}
\nonumber\\
&\hspace{-2cm} \times
 \delta(p_2^+-x_2P_A^+)\delta(p_3^+ - k^+ - x_3 P_A^+)\delta^{(d-2)}(p_{3\perp}-k_{3\perp}-k_\perp) \textrm{Tr}\Big[(i\gamma_\perp^\nu\gamma^+)
(-i)
\nonumber\\
&\hspace{-2cm} \times (k_2^+g_{\nu\mu}-k_{2\nu}n_\mu)T^b(-ig_sT^a)
\frac{in^\rho}{k^++i\varepsilon}
 \mathcal {M}_A^{b,\mu}(p_2,p_3)   (ig_s\gamma_\rho T^a) \frac{-i( p_3-k)\cdot \gamma}{( p_3-k)^2-i\varepsilon}\Big]
 \nonumber\\
&\hspace{-2cm} \times (-2\pi)\delta(k^2)\theta(k^+),
\end{align}
which results in
\begin{align}
h(x_1,x_2,k_{3\perp})\Big\vert_{5k} =& \frac{i(x_2+y_1)}{4(d-2) (x_2+x_1)}\Big(C_F-\frac{C_A}{2}\Big) \Big(\frac{\mu^2e^{\gamma_E}}{4\pi}\Big)^\epsilon
 \int\frac{d^{d}k}{(2\pi)^d}\frac{1}{k_2^+}
 \textrm{Tr}\Big[(i\gamma_\perp^\nu\gamma^+)
\nonumber\\
&\hspace{-2cm} \times (-i)
 (k_2^+g_{\nu\mu}-\hat k_{2\nu}n_\mu)(-ig_s)
\frac{in^\rho}{k^++i\varepsilon}
 (i\gamma^\mu_\perp \gamma^-)   (ig_s\gamma_\rho ) \frac{-i( \hat p_3-k)\cdot \gamma}{( \hat p_3-k)^2-i\varepsilon}\Big]
 \nonumber\\
&\hspace{-2cm} \times (-2\pi)\delta(k^2)\theta(k^+) h(y_1,x_2,k_{3\perp}+k_\perp).
\end{align}
Here still only $h$ contributes. We integrate over $k^-$ and obtain
\begin{align}\label{eq:5k}
h(x_1,x_2,k_{3\perp})\Big\vert_{5k}
=& 2\alpha_s \Big(C_F-\frac{C_A}{2}\Big)
 \int_{x_1}^\infty dy_1 \Big(\frac{1}{y_1-x_1} -\frac{i\pi}{2}\delta(y_1-x_1)\Big)
 \nonumber\\
&\times
 \Big(\frac{\mu^2e^{\gamma_E}}{4\pi}\Big)^\epsilon \int\frac{d^{2-2\epsilon}k_\perp}{(2\pi)^{2-2\epsilon}}\frac{1}{k_T^2}
h(y_1,x_2,k_{3\perp}+k_\perp).
\end{align}
The result contains the rapidity divergence at $x_1=y_1$ and IR divergence at $k_\perp=0$.
According to \cite{Gehrmann:2014yya}
\begin{align}
\frac{\mu^{2\epsilon}}{\pi^{1-\epsilon}} \int\frac{d^{2-2\epsilon}k_\perp}{k_T^{2+2\delta}}e^{ik_\perp\cdot b_\perp}
=\mu^{2\epsilon}e^{-2(\epsilon+\delta)\gamma_E}\frac{\Gamma(-\epsilon-\delta)}{\Gamma(1+\delta)}
\Big(\frac{b_T^2}{4e^{-2\gamma_E}}\Big)^{\epsilon+\delta},
\end{align}
 we have
 \begin{subequations}\label{eq:yll}
 \begin{align}
\frac{(4\pi)^\epsilon \mu^{2\epsilon}}{\pi^{1-\epsilon}} \int\frac{d^{2-2\epsilon}k_\perp}{k_T^{2}}e^{ik_\perp\cdot b_\perp}
=&\Big(\frac{4\pi b_T^2\mu^2}{4}\Big)^{\epsilon}\Gamma(-\epsilon) ,\\
-\frac{(4\pi)^\epsilon\mu^{2\epsilon}}{\pi^{1-\epsilon}} \int\frac{d^{2-2\epsilon}k_\perp}{k_T^{2}}\ln k_T^2 e^{ik_\perp\cdot b_\perp}
=&\Big(\frac{4\pi b_T^2\mu^2}{4}\Big)^{\epsilon}\Gamma(-\epsilon)\Big(-\psi(-\epsilon)
+\ln\frac{b_T^2}{4}+\gamma_E \Big),
\end{align}
\end{subequations}
where $\psi(x)=\Gamma^\prime(x)/\Gamma(x)$.
Applying the above equations, we can convert Eq. \eqref{eq:5k} into a form in $b$-space,
\begin{align}\label{eq:5kb}
\tilde{h}(x_1,x_2,b_{\perp})\Big\vert_{5k}
=& 2\alpha_s \Big(C_F-\frac{C_A}{2}\Big)
 \int_{x_1}^\infty dy_1 \Big(\frac{1}{y_1-x_1} -\frac{i\pi}{2}\delta(y_1-x_1)\Big) \frac{1}{4\pi}
\Big(-\frac{1}{\epsilon_{\textrm{IR}}}-\ln\mu^2\Big)
 \nonumber\\
&\times
\tilde{h}(y_1,x_2,b_\perp)+\cdots.
\end{align}
Here $\cdots$ denotes the finite terms that are irrelevant.
Similarly, for Fig. \ref{fig:evo}(l)-(o) we can derive
\begin{subequations}\label{eq:fig5r}
\begin{align}
\tilde{h}(x_1,x_2,b_{\perp})\Big\vert_{5l}
=& 2\alpha_s \Big(C_F-\frac{C_A}{2}\Big)
 \int_{x_2}^\infty dy_2 \Big(\frac{-x_2}{(y_2-x_2)y_2} +\frac{i\pi}{2}\delta(y_2-x_2) \Big)
 \nonumber\\
&\times
 \frac{1}{4\pi}
\Big(-\frac{1}{\epsilon_{\textrm{IR}}}-\ln\mu^2\Big)
\tilde{h}(x_1,y_2,b_\perp)+\cdots,\\
\tilde{h}(x_1,x_2,b_{\perp})\Big\vert_{5m}
=& 2\alpha_s C_F
 \int_{x_2}^\infty dy_2 \Big( \frac{x_2}{(y_2-x_2)y_2} -\frac{i\pi}{2}\delta(y_2-x_2) \Big)
 \nonumber\\
&\times
 \frac{1}{4\pi}
\Big(-\frac{1}{\epsilon_{\textrm{IR}}}-\ln\mu^2\Big)
\tilde{h}(x_1,y_2,b_\perp)+\cdots,\\
\tilde{h}(x_1,x_2,b_{\perp})\Big\vert_{5n}
=& \alpha_s \frac{C_A}{2}
 \int_{x_2}^\infty dy_2 \Big( \frac{x_2(x_2+y_2)(x_1+y_2)}{(y_2-x_2)(x_1+x_2)y_2^2} +i\pi\delta(y_2-x_2)\Big)
 \nonumber\\
&\times
\frac{1}{4\pi}
\Big(-\frac{1}{\epsilon_{\textrm{IR}}}-\ln\mu^2\Big)
\tilde{h}(x_1,y_2,b_\perp)+\cdots,\\
\tilde{h}(x_1,x_2,b_{\perp})\Big\vert_{5o}
=& 2\alpha_s \Big(C_F-\frac{C_A}{2}\Big)
 \int_{x_1}^\infty dy_1 \Big( \frac{x_1(y_1+x_2)}{(y_1-x_1)y_1(x_1+x_2)} +\frac{i\pi}{2}\delta(y_1-x_1) \Big)
 \nonumber\\
&\times
 \frac{1}{4\pi}
\Big(-\frac{1}{\epsilon_{\textrm{IR}}}-\ln\mu^2\Big)
\tilde{h}(y_1,x_2,b_\perp)+\cdots.
\end{align}
\end{subequations}

Summing up the IR divergences in Eqs.\eqref{eq:5a}, \eqref{eq:fig5v}, \eqref{eq:5kb} and \eqref{eq:fig5r}, we have
\begin{align}\label{eq:IR}
&\tilde{h}(x_1,x_2,b_{\perp})\Big\vert_{\textrm{Fig.5, IR}}\nonumber\\
=&  \frac{2\alpha_s}{4\pi} \Big\{ \Big[ 2C_F + \frac{C_A}{2} + \Big( C_F-\frac{C_A}{2} \Big)\Big( - \ln\frac{x_1+x_2}{x_1}\Big) + \frac{C_A}{2}\Big( -\ln \frac{x_1+x_2}{x_2} \Big) \Big]\tilde{h}(x_1,x_2,b_{\perp})
\nonumber\\
& + \frac{C_A}{2} \Big[2\int_{x_2}^\infty dy_2 \Big[\frac{x_2}{(y_2-x_2)y_2}\Big]_+ + \int_{x_2}^\infty dy_2  \frac{x_2(y_2-x_1)}{2y_2^2(x_1+x_2)} \Big]\tilde{h}(x_1,y_2,b_{\perp})
\nonumber\\
& + \Big(C_F-\frac{C_A}{2}\Big) \Big[2\int_{x_1}^\infty dy_1 \Big[\frac{x_1}{(y_1-x_1)y_1}\Big]_+ + \int_{x_1}^\infty dy_1 \Big(\frac{2}{y_1} -\frac{x_2}{y_1(x_2+x_1)} \Big) \Big]\tilde{h}(y_1,x_2,b_{\perp})
\nonumber\\
& + \frac{C_A}{2} \Big[\int_{-\infty}^\infty dy_2 \Big( \theta(x_2-y_2) \frac{x_1( x_2 + y_2)}{(y_1+y_2) y_1y_2}
 + \theta(y_2-x_2)  \frac{x_2( x_2 + y_2)}{(y_1+y_2) y_2^2}   \Big)\tilde{h}(y_1,y_2,b_{\perp})
  \nonumber\\
& +\int_{x_1}^\infty dy_1 \frac{x_1}{(y_1-x_1)y_1} \Big(\tilde{h}(y_1,y_2,b_{\perp}) - \tilde{h}(x_1,x_2,b_{\perp})\Big) \nonumber\\
& + \int_{x_2}^\infty dy_2 \Big(\frac{x_2^2}{(y_2-x_2)y_2^2}\tilde{h}(y_1,y_2,b_{\perp})
-\frac{x_2}{(y_2-x_2)y_2}\tilde{h}(x_1,x_2,b_{\perp}) \Big) \Big] \nonumber\\
&+ \Big(C_F - \frac{C_A}{2} \Big) \int_{-\infty}^\infty dy_2 \Big[ \theta(y_2-x_1)\frac{-x_1( x_2-y_1)}{(y_1+y_2) y_2^2}
- \theta(x_1-y_2)\frac{x_2( x_2-y_1)}{(y_1+y_2) y_2y_1}  \Big]
\nonumber\\
&\times
\tilde{h}(y_1,y_2,b_{\perp}) +  C_F \int_{-\infty}^\infty dy_2 \frac{-x_1x_2}{(y_1+y_2)^2y_2} \tilde{h}(y_1,y_2,b_{\perp}) \Big\}
\Big( -\frac{1}{\epsilon_{\textrm{IR}}} -\ln\mu^2 \Big)
 .
\end{align}
To derive the above result, we have used the following relations
\begin{subequations}
\begin{align}
\int_{x_1}^{\infty}dy_1\frac{x_1}{y_1(y_1-x_1)}=\int_{0}^{1}dy\frac{1}{1-y}+\ln x_1=\int_{0}^{1}dy\frac{1}{y}+\ln x_1,\\
\int_{x_2}^{\infty}dy_2\frac{x_2}{y_2(y_2-x_2)}=\int_{0}^{1}dy\frac{1}{1-y}+\ln x_2=\int_{0}^{1}dy\frac{1}{y}+\ln x_2.
\end{align}
\end{subequations}
In Eq. \eqref{eq:IR}, we find the rapidity divergences
associated with $1/\epsilon_{\textrm{IR}}$ are canceled. After this cancellation, the plus-distribution appears, which is defined as
\begin{align}
\int_x^\infty dy [f(y-x)]_+g(y)= \int_x^\infty dy f(y-x)\Big(g(y)-g(x)\Big).
\end{align}
In addition, we find the imaginary parts are also canceled.

We now derive the total $1/\epsilon$ divergences and rapidity divergences of the diagrams in Fig. \ref{fig:evo}. The contributions from the virtual diagrams vanish due to the cancellation between $1/\epsilon_{\textrm{UV}}$ and $1/\epsilon_{\textrm{IR}}$. Therefore, we only need to consider the real diagrams in Fig. \ref{fig:evo}(k)-(o). To regularize the rapidity divergences, we use the exponential regulator. This regulator is introduced by multiplying the phase space measure in Eq.\eqref{eq:AMk} by an exponential factor:
\begin{align}
\int \frac{d^dk}{(2\pi)^d}(2\pi)\delta(k^2) \to \lim_{\tau\to 0} \int \frac{d^dk}{(2\pi)^d}(2\pi)\delta(k^2) \exp(-b_0\tau (k^++k^-)),
\end{align}
with $b_0=2e^{-\gamma_E}$. For Fig. \ref{fig:evo}(k), the exponential factor can be rewritten as
\begin{align}
 \exp(-b_0\tau (k^++k^-))= \exp\Big[-b_0\tau \Big((y_1-x_1)P_A^+ + \frac{k_T^2}{2(y_1-x_1)P_A^+} \Big)\Big].
\end{align}
We can drop the term proportional to $\tau (y_1-x_1)$, as it has no contribution in the limit $\tau\to 0$. We then have
\begin{align}
h(x_1,x_2,k_{3\perp})\Big\vert_{5k}
=& 2\alpha_s \Big(C_F-\frac{C_A}{2}\Big) \Big(\frac{\mu^2e^{\gamma_E}}{4\pi}\Big)^\epsilon \int\frac{d^{2-2\epsilon}k_\perp}{(2\pi)^{2-2\epsilon}}\frac{1}{k_T^2} \int dy_1 \Big[ \frac{1}{y_1-x_1}
 \nonumber\\
&\times \exp\Big[ \frac{-b_0\tau k_T^2}{2(y_1-x_1)P_A^+} \Big] -\frac{i\pi}{2}\delta(y_1-x_1)\Big]
h(y_1,x_2,k_{3\perp}+k_\perp).
\end{align}
The term in the exponent provides the service of regularizing the rapidity divergence. Using
\begin{align}
\frac{x_1}{(y_1-x_1)y_1} \exp\Big[ \frac{-b_0\tau k_T^2}{2(y_1-x_1)P_A^+} \Big]
=&-\ln\Big(\frac{\tau k_T^2}{x_1P_A^+}\Big)\delta(y_1-x_1)+ \Big[\frac{x_1}{(y_1-x_1)y_1}\Big]_+ + \mathcal {O}(\tau),
\end{align}
we obtain
\begin{align}
h(x_1,x_2,k_{3\perp})\Big\vert_{5k}
=& 2\alpha_s \Big(C_F-\frac{C_A}{2}\Big)\Big(\frac{\mu^2e^{\gamma_E}}{4\pi}\Big)^\epsilon \int\frac{d^{2-2\epsilon}k_\perp}{(2\pi)^{2-2\epsilon}}\frac{1}{k_T^2}
 \Big\{\int_{x_1}^\infty dy_1 \Big(-\ln\frac{\tau k_T^2}{x_1 P_A^+}
 \nonumber\\&\times \delta(y_1-x_1)
  +\Big[\frac{x_1}{(y_1-x_1)y_1}\Big]_+ \Big)  + \int_{x_1}^\infty dy_1 \Big(\frac{1}{y_1} -\frac{i\pi}{2}\delta(y_1-x_1) \Big) \Big\}
\nonumber\\&\times
h(y_1,x_2,k_{3\perp}+k_\perp).
\end{align}
We find the rapidity divergences are regulated by regulator $\tau$.
Applying Eq.\eqref{eq:yll}, we derive
\begin{align}\label{eq:5kR}
\tilde{h}(x_1,x_2,b_{\perp})\Big\vert_{5k}
=& 2\alpha_s \Big(C_F-\frac{C_A}{2}\Big) \Big\{\int_{x_1}^\infty dy_1 \Big[-\ln\frac{\tau }{x_1 P_A^+}\delta(y_1-x_1)+\Big[\frac{x_1}{(y_1-x_1)y_1}\Big]_+  \nonumber\\
& + \frac{1}{y_1} -\frac{i\pi}{2}\delta(y_1-x_1)  \Big] \frac{1}{4\pi} \Big(-\frac{1}{\epsilon}-\ln\frac{ b_T^2\mu^2}{4e^{-2\gamma_E}}\Big)
\nonumber\\
&+ \int_{x_1}^\infty dy_1 \delta(y_1-x_1)\frac{1}{4\pi}  \Big(\frac{1}{\epsilon^2}+\frac{\ln\mu^2}{\epsilon}
+\frac{1}{2}\ln^2\mu^2 + \cdots \Big)\Big\}
\nonumber\\
&\times
\tilde{h}(y_1,x_2,b_{\perp}).
\end{align}
Again, the $\cdots$ represents finite terms that are not relevant to our study.
Similarly, we can derive
\begin{subequations}\label{eq:5R}
\begin{align}
\tilde{h}(x_1,x_2,b_{\perp})\Big\vert_{5l}
=& - 2\alpha_s \Big(C_F-\frac{C_A}{2}\Big) \Big\{\int_{x_2}^\infty dy_2 \Big[-\ln\frac{\tau }{x_2 P_A^+}\delta(y_2-x_2) +\Big[\frac{x_2}{(y_2-x_2)y_2}\Big]_+
\nonumber\\
&- \frac{i\pi}{2} \delta(y_2-x_2)    \Big]\frac{1}{4\pi}\Big(-\frac{1}{\epsilon}-\ln\frac{ b_T^2\mu^2}{4e^{-2\gamma_E}}\Big)
\nonumber\\
&+ \int_{x_2}^\infty dy_2 \delta(y_2-x_2)\frac{1}{4\pi}  \Big(\frac{1}{\epsilon^2}+\frac{\ln\mu^2}{\epsilon}
+\frac{1}{2}\ln^2\mu^2 + \cdots\Big)\Big\}
\nonumber\\
&\times\tilde{h}(x_1,y_2,b_{\perp}),\\
\tilde{h}(x_1,x_2,b_{\perp})\Big\vert_{5m}
=&  2\alpha_s C_F \Big\{\int_{x_2}^\infty dy_2 \Big[-\ln\frac{\tau }{x_2 P_A^+}\delta(y_2-x_2) +\Big[\frac{x_2}{(y_2-x_2)y_2}\Big]_+
\nonumber\\
&- \frac{i\pi}{2} \delta(y_2-x_2)    \Big]\frac{1}{4\pi}\Big(-\frac{1}{\epsilon}-\ln\frac{ b_T^2\mu^2}{4e^{-2\gamma_E}}\Big)
\nonumber\\
&+ \int_{x_2}^\infty dy_2 \delta(y_2-x_2)\frac{1}{4\pi}  \Big(\frac{1}{\epsilon^2}+\frac{\ln\mu^2}{\epsilon}
+\frac{1}{2}\ln^2\mu^2 + \cdots\Big)\Big\}
\nonumber\\
&\times\tilde{h}(x_1,y_2,b_{\perp}),\\
\tilde{h}(x_1,x_2,b_{\perp})\Big\vert_{5n}
=&  2\alpha_s \Big(\frac{C_A}{2}\Big) \Big\{\int_{x_2}^\infty dy_2 \Big[-\ln\frac{\tau }{x_2 P_A^+}\delta(y_2-x_2) +\Big[\frac{x_2}{(y_2-x_2)y_2}\Big]_+
\nonumber\\
&+ \frac{x_2(y_2-x_1)}{2y_2^2(x_1+x_2)} + \frac{i\pi}{2} \delta(y_2-x_2)    \Big]\frac{1}{4\pi}\Big(-\frac{1}{\epsilon}-\ln\frac{ b_T^2\mu^2}{4e^{-2\gamma_E}}\Big)
\nonumber\\
&+ \int_{x_2}^\infty dy_2 \delta(y_2-x_2)\frac{1}{4\pi}  \Big(\frac{1}{\epsilon^2}+\frac{\ln\mu^2}{\epsilon}
+\frac{1}{2}\ln^2\mu^2 + \cdots\Big)\Big\}
\nonumber\\
&\times\tilde{h}(x_1,y_2,b_{\perp}),\\
\tilde{h}(x_1,x_2,b_{\perp})\Big\vert_{5o}
=& 2\alpha_s \Big(C_F-\frac{C_A}{2}\Big) \Big\{\int_{x_1}^\infty dy_1 \Big[-\ln\frac{\tau }{x_1 P_A^+}\delta(y_1-x_1)+\Big[\frac{x_1}{(y_1-x_1)y_1}\Big]_+  \nonumber\\
& + \frac{x_1}{y_1(x_1+x_2)} +\frac{i\pi}{2}\delta(y_1-x_1)  \Big] \frac{1}{4\pi} \Big(-\frac{1}{\epsilon}-\ln\frac{ b_T^2\mu^2}{4e^{-2\gamma_E}}\Big)
\nonumber\\
&+ \int_{x_1}^\infty dy_1 \delta(y_1-x_1)\frac{1}{4\pi}  \Big(\frac{1}{\epsilon^2}+\frac{\ln\mu^2}{\epsilon}
+\frac{1}{2}\ln^2\mu^2 + \cdots \Big)\Big\}
\nonumber\\
&\times
\tilde{h}(y_1,x_2,b_{\perp}).
\end{align}
\end{subequations}
We can obtain the total $1/\epsilon$ divergences and rapidity divergences by summing the results in Eqs.\eqref{eq:5kR} and \eqref{eq:5R}. Using this summation to subtract the IR divergences in Eq.\eqref{eq:IR}, we then derive the UV divergences and the regularized rapidity divergences of the diagrams in Fig. \ref{fig:evo}. Combining the result of Fig. \ref{fig:self-energy} in Eq. \eqref{eq:SE}, we then derive
\begin{align}\label{eq:UV}
&\tilde{h}(x_1,x_2,b_{\perp})\Big\vert_{\textrm{UV}}
\nonumber\\
=& \frac{4\alpha_s}{4\pi}C_F\Big(\frac{1}{\epsilon^2}+\frac{\ln\mu^2}{\epsilon}
+\frac{1}{2}\ln^2\mu^2\Big) \tilde{h}(x_1,x_2,b_{\perp})
 \nonumber\\
 &+  \frac{2\alpha_s}{4\pi} \Big\{ \Big[\frac{3}{2}C_F + \Big( C_F-\frac{C_A}{2} \Big)\Big( - \ln\frac{x_1+x_2}{x_1}\Big) + \frac{C_A}{2}\Big( -\ln \frac{x_1+x_2}{x_2} \Big) \Big]\tilde{h}(x_1,x_2,b_{\perp})
\nonumber\\
& + \frac{C_A}{2} \Big[\int_{-\infty}^\infty dy_2 \Big[ \theta(x_2-y_2) \frac{x_1( x_2 + y_2)}{(y_1+y_2) y_1y_2}
 + \theta(y_2-x_2)  \frac{x_2( x_2 + y_2)}{(y_1+y_2) y_2^2}   \Big]\tilde{h}(y_1,y_2,b_{\perp})
  \nonumber\\
& +\int_{x_1}^\infty dy_1 \frac{x_1}{(y_1-x_1)y_1} \Big(\tilde{h}(y_1,y_2,b_{\perp}) - \tilde{h}(x_1,x_2,b_{\perp})\Big) \nonumber\\
& + \int_{x_2}^\infty dy_2 \Big(\frac{x_2^2}{(y_2-x_2)y_2^2}\tilde{h}(y_1,y_2,b_{\perp})
-\frac{x_2}{(y_2-x_2)y_2}\tilde{h}(x_1,x_2,b_{\perp}) \Big) \Big] \nonumber\\
&+ \Big(C_F - \frac{C_A}{2} \Big) \int_{-\infty}^\infty dy_2 \Big[ \theta(y_2-x_1)\frac{-x_1( x_2-y_1)}{(y_1+y_2) y_2^2}
- \theta(x_1-y_2)\frac{x_2( x_2-y_1)}{(y_1+y_2) y_2y_1}  \Big]
\nonumber\\
&\times
\tilde{h}(y_1,y_2,b_{\perp}) +  C_F \int_{-\infty}^\infty dy_2 \frac{-x_1x_2}{(y_1+y_2)^2y_2} \tilde{h}(y_1,y_2,b_{\perp}) \Big\}
\Big( \frac{1}{\epsilon} +\ln\mu^2 \Big)
\nonumber\\
&
 +  \frac{2\alpha_s}{4\pi} \Big[\frac{C_A}{2} \int_{x_2}^\infty dy_2  \Big(2\ln\frac{\tau }{x_2 P_A^+}\delta(y_2-x_2)\Big) \tilde{h}(x_1,y_2,b_{\perp})
 + \Big(C_F-\frac{C_A}{2}\Big)
  \nonumber\\
&\times \int_{x_1}^\infty dy_1 \Big(2\ln\frac{\tau }{x_1 P_A^+}\delta(y_1-x_1) \Big) \tilde{h}(y_1,x_2,b_{\perp}) \Big]
 \Big(\frac{1}{\epsilon}+\ln\frac{ b_T^2\mu^2}{4e^{-2\gamma_E}}\Big)
 .
\end{align}
With the above result, we now can derive the evolution equations for $\tilde{h}$. The RG evolution equation is given by
\begin{align}
\frac{d}{d\ln\mu^2}\tilde{h}(x_1,x_2,b_{\perp})
=&
\Big[\frac{ \alpha_sC_F}{\pi}   \ln\frac{\mu^2 }{(x_1+x_2)\nu P_A^+}+ \mathbb{K}_A \Big]\tilde{h}(x_1,x_2,b_{\perp}),
\end{align}
where $\nu=1/\tau$, and $\mathbb{K}_A$ is the integral kernel that acts on the TMD PDF, defined by
\begin{align}\label{eq:KA}
&\mathbb{K}_A \tilde{h}(x_1,x_2,b_{\perp})
\nonumber\\
=& \frac{ \alpha_s}{2\pi} \Big\{ \Big[\frac{3}{2}C_F + \Big( C_F-\frac{C_A}{2} \Big)\ln\frac{x_1+x_2}{x_1} + \frac{C_A}{2}\ln \frac{x_1+x_2}{x_2} \Big]\tilde{h}(x_1,x_2,b_{\perp})
\nonumber\\
& + \frac{C_A}{2} \Big[\int_{-\infty}^\infty dy_2 \Big[ \theta(x_2-y_2) \frac{x_1( x_1 + 2x_2)}{(x_1+x_2)^2 y_1}
 + \theta(y_2-x_2)  \frac{x_2^2(x_1+ x_2 + y_2)}{(x_1+x_2)^2 y_2^2}   \Big]\tilde{h}(y_1,y_2,b_{\perp})
  \nonumber\\
& +\int_{x_1}^\infty dy_1 \frac{x_1}{(y_1-x_1)y_1} \Big(\tilde{h}(y_1,y_2,b_{\perp}) - \tilde{h}(x_1,x_2,b_{\perp})\Big) \nonumber\\
& + \int_{x_2}^\infty dy_2 \Big(\frac{x_2^2}{(y_2-x_2)y_2^2}\tilde{h}(y_1,y_2,b_{\perp})
-\frac{x_2}{(y_2-x_2)y_2}\tilde{h}(x_1,x_2,b_{\perp}) \Big) \Big] \nonumber\\
&+ \Big(C_F - \frac{C_A}{2} \Big) \int_{-\infty}^\infty dy_2 \Big[ \theta(y_2-x_1)\frac{x_1(x_1^2+x_1(x_2-y_2)-2 x_2y_2)}{(x_1+x_2)^2 y_2^2}
\nonumber\\&
- \theta(x_1-y_2)\frac{x_2^2}{(x_1+x_2)^2 y_1}  \Big]
\tilde{h}(y_1,y_2,b_{\perp})  \Big\}
 .
\end{align}
We find our evolution kernel $\mathbb{K}_A$ agrees with the kernel $\Upsilon_{x_1x_2x_3}+2\mathbb{P}^A_{x_2x_1}$ in Ref.\cite{Rodini:2023plb}, but differs from the kernel $\Upsilon_{x_1x_2x_3}+2\mathbb{P}^A_{x_2x_1}$ in Ref.\cite{Rodini:2022wki} by a sign \footnote{We have corresponded with the authors of Refs. \cite{Rodini:2022wki,Rodini:2023plb} and verified that Ref. \cite{Rodini:2022wki} contains a typographical error, while the corresponding expressions in Ref. \cite{Rodini:2023plb} are correct.}.
The evolution equation with respect to the rapidity scale $\nu$ is given by
\begin{align}
\frac{d}{d\ln\nu}\tilde{h}(x_1,x_2,b_{\perp})=& -\frac{\alpha_s C_F}{\pi}\ln\frac{ b_T^2\mu^2}{4e^{-2\gamma_E}} \tilde{h}(x_1,x_2,b_{\perp}).
\end{align}

Following the same route, we derive the evolution equations for distributions $f^\perp-ig^\perp$, $f^\perp+ig^\perp$ and $h^\perp$. The RG evolution equations read as
\begin{subequations}\label{eq:RG-Unsub}
\begin{align}
\frac{d}{d\ln\mu^2}(\tilde{f}^{\perp(1)}-i\tilde{g}^{\perp(1)})
=&
\Big[\frac{ \alpha_sC_F}{\pi}   \ln\frac{\mu^2 }{(x_1+x_2)\nu P_A^+}+ \mathbb{K}_A \Big](\tilde{f}^{\perp(1)}-i\tilde{g}^{\perp(1)}),\\
\frac{d}{d\ln\mu^2}(\tilde{f}^{\perp(1)}+i\tilde{g}^{\perp(1)})
=&
\Big[\frac{ \alpha_sC_F}{\pi}   \ln\frac{\mu^2 }{(x_1+x_2)\nu P_A^+}+ \mathbb{K}_B \Big](\tilde{f}^{\perp(1)}+i\tilde{g}^{\perp(1)}),\\
\frac{d}{d\ln\mu^2}\tilde{h}^{\perp(2)}
=&
\Big[\frac{ \alpha_sC_F}{\pi}   \ln\frac{\mu^2 }{(x_1+x_2)\nu P_A^+}+ \mathbb{K}_B \Big]\tilde{h}^{\perp(2)},
\end{align}
\end{subequations}
here the argument of distributions $(x_1,x_2,b_{\perp})$ is omitted. The evolution kernel $\mathbb{K}_B$ is given by
\begin{align}\label{eq:KB}
&\mathbb{K}_B \tilde{h}^{\perp(2)}(x_1,x_2,b_{\perp})
\nonumber\\
=& \frac{ \alpha_s}{2\pi} \Big\{ \Big[\frac{3}{2}C_F + \Big( C_F-\frac{C_A}{2} \Big)\ln\frac{x_1+x_2}{x_1} + \frac{C_A}{2}\ln \frac{x_1+x_2}{x_2} \Big]\tilde{h}^{\perp(2)}(x_1,x_2,b_{\perp})
\nonumber\\
& + \frac{C_A}{2} \Big[
\int_{x_1}^\infty dy_1 \frac{x_1}{(y_1-x_1)y_1} \Big(\tilde{h}^{\perp(2)}(y_1,y_2,b_{\perp}) - \tilde{h}^{\perp(2)}(x_1,x_2,b_{\perp})\Big) \nonumber\\
& + \int_{x_2}^\infty dy_2 \Big(\frac{x_2^2}{(y_2-x_2)y_2^2}\tilde{h}^{\perp(2)}(y_1,y_2,b_{\perp})
-\frac{x_2}{(y_2-x_2)y_2}\tilde{h}^{\perp(2)}(x_1,x_2,b_{\perp}) \Big) \Big] \nonumber\\
&+ \Big(C_F - \frac{C_A}{2} \Big) \int_{-\infty}^\infty dy_2  \theta(y_2-x_1)\frac{x_1}{ y_2^2}
\tilde{h}^{\perp(2)}(y_1,y_2,b_{\perp})  \Big\}.
\end{align}
Again, our evolution kernel $\mathbb{K}_B$ differs from the the kernel $\Upsilon_{x_1x_2x_3}+2\mathbb{P}^B_{x_2x_1}$ in Ref.\cite{Rodini:2022wki} by a sign.
The evolution in the rapidity scale $\nu$ is the same for all distributions. It reads
\begin{align}\label{eq:CS-Unsub}
\frac{d}{d\ln\nu}H(x_1,x_2,b_{\perp})=& -\frac{\alpha_s C_F}{\pi}\ln\frac{ b_T^2\mu^2}{4e^{-2\gamma_E}} H(x_1,x_2,b_{\perp}),
\end{align}
here $H\in \{\tilde{f}^{\perp(1)}-i\tilde{g}^{\perp(1)}, \tilde{f}^{\perp(1)}+i\tilde{g}^{\perp(1)}, \tilde{h}^{\perp(2)}\}$.

\section{Soft subtraction}\label{sec:sub}

For TMD factorization, it is well known that a soft subtraction \cite{Collins:2011zzd,Echevarria:2011epo,Chiu:2012ir,Chen:2016hgw} or a zero-bin subtraction \cite{Manohar:2006nz} is necessary to avoid double-counting between the collinear sectors and the soft sector. To eliminate double-counting,
we divide each TMD PDF by a soft factor. It has been demonstrated in \cite{Vladimirov:2021hdn} that the soft factor is identical for both the twist-2 and twist-3 TMD PDFs, and is given by the LP soft function
\begin{align}
\mathcal {S}_{\textrm{LP}}(b_{\perp})=\frac{1}{N_c} \textrm{Tr}\langle0| {\mathcal L}_l ( b_\perp)  {\mathcal L}^\dag_n ( b_\perp){\mathcal L}_n ( 0){\mathcal L}_l^\dag ( 0)   |0\rangle.
\end{align}
We then implement the soft subtraction for twist-3 TMD PDFs by making the following replacements:
\begin{align}
H(x_1,x_2,b_{\perp})\to \frac{H(x_1,x_2,b_{\perp})}{\sqrt{\mathcal {S}_{\textrm{LP}}(b_{\perp})}}=H^{\textrm{sub}}(x_1,x_2,b_{\perp}),
\end{align}
where $H\in \{\tilde{f}^{\perp(1)}-i\tilde{g}^{\perp(1)}, \tilde{f}^{\perp(1)}+i\tilde{g}^{\perp(1)},\tilde{h},\tilde{h}^{\perp(2)}\}$. The exponentially regularized LP soft function has been calculated to $\rm{N^3LO}$ in \cite{Li:2016axz}. According to their result we have
\begin{subequations}\label{eq:RG-S}
\begin{align}
\frac{d}{d\ln\mu^2}\mathcal {S}_{\textrm{LP}}(b_{\perp})=& \Big[\frac{\alpha_s C_F}{\pi}\ln\frac{ \mu^2}{\nu^2}+\mathcal {O}(\alpha_s ^2)\Big] \mathcal {S}_{\textrm{LP}}(b_{\perp}),\\
\frac{d}{d\ln\nu}\mathcal {S}_{\textrm{LP}}(b_{\perp})=& \Big[-\frac{2\alpha_s C_F}{\pi}\ln\frac{ b_T^2\mu^2}{4e^{-2\gamma_E}} +\mathcal {O}(\alpha_s ^2)\Big]\mathcal {S}_{\textrm{LP}}(b_{\perp}).
\end{align}
\end{subequations}
Based on Eqs. \eqref{eq:RG-Unsub}, \eqref{eq:CS-Unsub} and \eqref{eq:RG-S}, we immediately obtain the evolution equations for the subtracted distributions $H^{\textrm{sub}}(x_1,x_2,b_{\perp})$. The RG evolution equations read as
\begin{subequations}\label{eq:RG-sub}
\begin{align}
\frac{d}{d\ln\mu^2}(\tilde{f}^{\perp(1),\textrm{sub}}-i\tilde{g}^{\perp(1),\textrm{sub}})
=&
\Big[\frac{ \alpha_sC_F}{2\pi}   \ln\frac{\mu^2 }{((x_1+x_2) P_A^+)^2}+ \mathbb{K}_A \Big](\tilde{f}^{\perp(1),\textrm{sub}}-i\tilde{g}^{\perp(1),\textrm{sub}}),\\
\frac{d}{d\ln\mu^2}(\tilde{f}^{\perp(1),\textrm{sub}}+i\tilde{g}^{\perp(1),\textrm{sub}})
=&
\Big[\frac{ \alpha_sC_F}{2\pi}   \ln\frac{\mu^2 }{((x_1+x_2) P_A^+)^2}+ \mathbb{K}_B \Big](\tilde{f}^{\perp(1),\textrm{sub}}+i\tilde{g}^{\perp(1),\textrm{sub}}),\\
\frac{d}{d\ln\mu^2}\tilde{h}^{\textrm{sub}}
=&
\Big[\frac{ \alpha_sC_F}{2\pi}   \ln\frac{\mu^2 }{((x_1+x_2) P_A^+)^2}+ \mathbb{K}_A \Big]\tilde{h}^{\textrm{sub}},\\
\frac{d}{d\ln\mu^2}\tilde{h}^{\perp(2),\textrm{sub}}
=&
\Big[\frac{ \alpha_sC_F}{2\pi}   \ln\frac{\mu^2 }{((x_1+x_2) P_A^+)^2}+ \mathbb{K}_B \Big]\tilde{h}^{\perp(2),\textrm{sub}}.
\end{align}
\end{subequations}
These results reproduce the RG evolution equations given in \cite{Rodini:2023plb} with $\zeta=((x_1+x_2) P_A^+)^2$. For the rapidity scale, we find the $\nu$ dependence vanished,
\begin{align}\label{eq:CS-sub-1}
\frac{d}{d\ln\nu}H^{\textrm{sub}}(x_1,x_2,b_{\perp})=& 0,
\end{align}
where $H\in \{\tilde{f}^{\perp(1)}-i\tilde{g}^{\perp(1)}, \tilde{f}^{\perp(1)}+i\tilde{g}^{\perp(1)}, \tilde{h},\tilde{h}^{\perp(2)}\}$. This indicates that the rapidity divergences cancel in the combination $H^{\textrm{sub}}=H/\sqrt{\mathcal {S}_{\textrm{LP}}}$.
However, this cancellation leaves a residual dependence on the scale $\zeta=((x_1+x_2) P_A^+)^2$, which reads as 
\begin{align}\label{eq:CS-sub-2}
\frac{d}{d\ln\zeta}H^{\textrm{sub}}(x_1,x_2,b_{\perp})=& -\frac{\alpha_s C_F}{2\pi}\ln\frac{ b_T^2\mu^2}{4e^{-2\gamma_E}} H^{\textrm{sub}}(x_1,x_2,b_{\perp}).
\end{align}
This evolution equation agrees with that in \cite{Rodini:2022wki,Rodini:2023plb}.

\section{Summary}\label{sec:summary}

In this paper, we calculated the RG evolution equations and the rapidity
scale evolution equations for four unpolarized twist-3 TMD PDFs $f^\perp-ig^\perp$, $f^\perp+ig^\perp$, $h$ and $h^\perp$ at $\mathcal {O}(\alpha_s)$. Compared to the studies in Refs. \cite{Vladimirov:2021hdn,Rodini:2022wki,Rodini:2023plb}, which are based on the background field method, we derived the evolution equations straightforwardly through diagram expansion. Additionally, instead of using the $\delta$ regulator, we employed the exponential regulator to regularize the rapidity divergences.

We began by separating the correlation functions at different twists using the equation of motion. Our calculations then showed that the contributions to the evolution from the twist-2 quark-quark correlation function cancel out. As a result, the twist-3 TMD PDFs will not mix with the twist-2 TMD PDFs during the evolution process. In our calculation, we first derived the IR divergences of all relevant diagrams, where the rapidity divergences associated with the IR divergences cancel between the virtual and real corrections. We then derived the total $1/\epsilon$ divergences and rapidity divergences of these diagrams, with the rapidity divergences regularized using the exponential regulator. Finally, by subtracting the IR divergences from the sum of $1/\epsilon$ divergences and rapidity divergences, we obtained the UV divergences and the regularized rapidity divergences. Based on this, we derived the RG evolution equations and the rapidity scale evolution equations for the twist-3 TMD PDFs, which are governed by eight homogeneous equations. After performing the soft subtraction, the rapidity divergences cancel between the TMD PDFs and soft factor, leaving a dependence on $\zeta$ in the subtracted TMD PDFs. We further derived the evolution equations for the subtracted twist-3 TMD PDFs. Our evolution kernels agree with the results in Ref.\cite{Rodini:2023plb}, but differ by a sign from those in Ref.\cite{Rodini:2022wki}.

Here, we have only considered the unpolarized twist-3 TMD PDFs, but our method can also be extended to calculate the evolution equations for polarized twist-3 TMD PDFs and TMD FFs. Additionally, given the advantages of the exponential regulator, using it could simplify the calculations for twist-3 TMD PDFs and TMD FFs. Our results can be applied to calculate twist-3 TMD PDFs at NNLO ($\mathcal{O}(\alpha_s^2)$). We leave these for future study.

\section{Acknowledgments}

We thank Jian-Ping Ma, Guang-Peng Zhang, Shuai Zhao and Xu-Chang Zheng for many useful discussions. We also thank S. Rodini and A. Vladimirov for discussing the results presented in their paper with us. This work is supported by the National Natural Science Foundation of China (Grant No. 12205124) and Jiangxi Provincial Natural Science Foundation (Grant No. 20242BAB20034).

\appendix
\section{The evolution equations in other support domains}
    \label{ap:expression}
In this appendix, we present the evolution equations in other support domains of $x_i$, $(i=1,2,3)$. The evolution equations with respect
to the scale $\mu$ have following formulas
\begin{align}
\frac{d}{d\ln\mu^2}H(x_1,x_2,b_{\perp})
=&
\Big[\frac{ \alpha_sC_F}{2\pi}   \ln\frac{\mu^2 }{((x_1+x_2) P_A^+)^2}+ \mathbb{K}_A \Big]H(x_1,x_2,b_{\perp}),
\end{align}
for $H\in \{ \tilde{f}^{\perp(1),\textrm{sub}}-i\tilde{g}^{\perp(1),\textrm{sub}},
\tilde{h}^{\textrm{sub}} \}$. And
\begin{align}
\frac{d}{d\ln\mu^2}H(x_1,x_2,b_{\perp})
=&
\Big[\frac{ \alpha_sC_F}{2\pi}   \ln\frac{\mu^2 }{((x_1+x_2) P_A^+)^2}+ \mathbb{K}_B \Big]H(x_1,x_2,b_{\perp}),
\end{align}
for $H\in \{ \tilde{f}^{\perp(1),\textrm{sub}}+i\tilde{g}^{\perp(1),\textrm{sub}},
\tilde{h}^{\perp(2),\textrm{sub}} \}$. The evolution kernels $\mathbb{K}_A$ and $\mathbb{K}_B$ are depend on the domain of $x_i$. The results for $x_1>0, x_2>0, x_3=x_1+x_2>0$ are given in Eq. \eqref{eq:KA} and Eq. \eqref{eq:KB}. The results for other domains are given as follows.

1. For $x_1>0, x_2<0, x_3>0$, we have
\begin{align}
&\mathbb{K}_A H(x_1,x_2,b_{\perp})
\nonumber\\
=& \frac{ \alpha_s}{2\pi} \Big\{ \Big[\frac{3}{2}C_F + \Big( C_F-\frac{C_A}{2} \Big)\ln\frac{|x_1+x_2|}{|x_1|} + \frac{C_A}{2}\ln \frac{|x_1+x_2|}{|x_2|} +\frac{C_A}{2}i\pi \Big]H(x_1,x_2,b_{\perp})
\nonumber\\
& + \frac{C_A}{2} \Big[\int_{-\infty}^\infty dy_2 \Big[ \theta(x_2-y_2) \frac{x_1( x_1 + 2x_2)}{(x_1+x_2)^2 y_1}
 - \theta(x_2-y_2)  \frac{x_2^2(x_1+ x_2 + y_2)}{(x_1+x_2)^2 y_2^2}   \Big]H(y_1,y_2,b_{\perp})
  \nonumber\\
& +\int_{x_1}^\infty dy_1 \frac{x_1}{(y_1-x_1)y_1} \Big(H(y_1,y_2,b_{\perp}) - H(x_1,x_2,b_{\perp})\Big) \nonumber\\
& + \int_{-\infty}^{x_2} dy_2 \Big(\frac{x_2^2}{(x_2-y_2)y_2^2}H(y_1,y_2,b_{\perp})
-\frac{x_2}{(x_2-y_2)y_2}H(x_1,x_2,b_{\perp}) \Big) \Big] \nonumber\\
&+ \Big(C_F - \frac{C_A}{2} \Big) \int_{-\infty}^\infty dy_2 \Big[ \theta(y_2-x_1)\frac{x_1(x_1^2+x_1(x_2-y_2)-2 x_2y_2)}{(x_1+x_2)^2 y_2^2}
\nonumber\\&
+ \theta(y_2-x_1)\frac{x_2^2}{(x_1+x_2)^2 y_1}  \Big]
H(y_1,y_2,b_{\perp})  \Big\}
 .
\end{align}
And
\begin{align}
&\mathbb{K}_B H(x_1,x_2,b_{\perp})
\nonumber\\
=& \frac{ \alpha_s}{2\pi} \Big\{ \Big[\frac{3}{2}C_F + \Big( C_F-\frac{C_A}{2} \Big)\ln\frac{|x_1+x_2|}{|x_1|} + \frac{C_A}{2}\ln \frac{|x_1+x_2|}{|x_2|} +\frac{C_A}{2}i\pi \Big]H(x_1,x_2,b_{\perp})
\nonumber\\
& + \frac{C_A}{2} \Big[
\int_{x_1}^\infty dy_1 \frac{x_1}{(y_1-x_1)y_1} \Big(H(y_1,y_2,b_{\perp}) - H(x_1,x_2,b_{\perp})\Big) \nonumber\\
& + \int^{x_2}_{-\infty} dy_2 \Big(\frac{x_2^2}{(x_2-y_2)y_2^2}H(y_1,y_2,b_{\perp})
-\frac{x_2}{(x_2-y_2)y_2}H(x_1,x_2,b_{\perp}) \Big) \Big] \nonumber\\
&+ \Big(C_F - \frac{C_A}{2} \Big) \int_{-\infty}^\infty dy_2  \theta(y_2-x_1)\frac{x_1}{ y_2^2}
H(y_1,y_2,b_{\perp})  \Big\}.
\end{align}

2. For $x_1>0, x_2<0, x_3<0$, we have
\begin{align}
&\mathbb{K}_A H(x_1,x_2,b_{\perp})
\nonumber\\
=& \frac{ \alpha_s}{2\pi} \Big\{ \Big[\frac{3}{2}C_F + \Big( C_F-\frac{C_A}{2} \Big)\ln\frac{|x_1+x_2|}{|x_1|} + \frac{C_A}{2}\ln \frac{|x_1+x_2|}{|x_2|} -\Big(C_F-\frac{C_A}{2}\Big)i\pi \Big]H(x_1,x_2,b_{\perp})
\nonumber\\
& + \frac{C_A}{2} \Big[\int_{-\infty}^\infty dy_2 \Big[ \theta(x_2-y_2) \frac{x_1( x_1 + 2x_2)}{(x_1+x_2)^2 y_1}
 - \theta(x_2-y_2)  \frac{x_2^2(x_1+ x_2 + y_2)}{(x_1+x_2)^2 y_2^2}   \Big]H(y_1,y_2,b_{\perp})
  \nonumber\\
& +\int_{x_1}^\infty dy_1 \frac{x_1}{(y_1-x_1)y_1} \Big(H(y_1,y_2,b_{\perp}) - H(x_1,x_2,b_{\perp})\Big) \nonumber\\
& + \int_{-\infty}^{x_2} dy_2 \Big(\frac{x_2^2}{(x_2-y_2)y_2^2}H(y_1,y_2,b_{\perp})
-\frac{x_2}{(x_2-y_2)y_2}H(x_1,x_2,b_{\perp}) \Big) \Big] \nonumber\\
&+ \Big(C_F - \frac{C_A}{2} \Big) \int_{-\infty}^\infty dy_2 \Big[ \theta(y_2-x_1)\frac{x_1(x_1^2+x_1(x_2-y_2)-2 x_2y_2)}{(x_1+x_2)^2 y_2^2}
\nonumber\\&
+ \theta(y_2-x_1)\frac{x_2^2}{(x_1+x_2)^2 y_1}  \Big]
H(y_1,y_2,b_{\perp})  \Big\}
 .
\end{align}
And
\begin{align}
&\mathbb{K}_B H(x_1,x_2,b_{\perp})
\nonumber\\
=& \frac{ \alpha_s}{2\pi} \Big\{ \Big[\frac{3}{2}C_F + \Big( C_F-\frac{C_A}{2} \Big)\ln\frac{|x_1+x_2|}{|x_1|} + \frac{C_A}{2}\ln \frac{|x_1+x_2|}{|x_2|} -\Big(C_F-\frac{C_A}{2}\Big)i\pi \Big]H(x_1,x_2,b_{\perp})
\nonumber\\
& + \frac{C_A}{2} \Big[
\int_{x_1}^\infty dy_1 \frac{x_1}{(y_1-x_1)y_1} \Big(H(y_1,y_2,b_{\perp}) - H(x_1,x_2,b_{\perp})\Big) \nonumber\\
& + \int^{x_2}_{-\infty} dy_2 \Big(\frac{x_2^2}{(x_2-y_2)y_2^2}H(y_1,y_2,b_{\perp})
-\frac{x_2}{(x_2-y_2)y_2}H(x_1,x_2,b_{\perp}) \Big) \Big] \nonumber\\
&+ \Big(C_F - \frac{C_A}{2} \Big) \int_{-\infty}^\infty dy_2  \theta(y_2-x_1)\frac{x_1}{ y_2^2}
H(y_1,y_2,b_{\perp})  \Big\}.
\end{align}

3. For $x_1<0, x_2>0, x_3>0$, we have
\begin{align}
&\mathbb{K}_A H(x_1,x_2,b_{\perp})
\nonumber\\
=& \frac{ \alpha_s}{2\pi} \Big\{ \Big[\frac{3}{2}C_F + \Big( C_F-\frac{C_A}{2} \Big)\ln\frac{|x_1+x_2|}{|x_1|} + \frac{C_A}{2}\ln \frac{|x_1+x_2|}{|x_2|} +\Big(C_F-\frac{C_A}{2}\Big)i\pi \Big]H(x_1,x_2,b_{\perp})
\nonumber\\
& + \frac{C_A}{2} \Big[\int_{-\infty}^\infty dy_2 \Big[ -\theta(y_2-x_2) \frac{x_1( x_1 + 2x_2)}{(x_1+x_2)^2 y_1}
 + \theta(y_2-x_2)  \frac{x_2^2(x_1+ x_2 + y_2)}{(x_1+x_2)^2 y_2^2}   \Big]H(y_1,y_2,b_{\perp})
  \nonumber\\
& +\int_{-\infty}^{x_1} dy_1 \frac{x_1}{(x_1-y_1)y_1} \Big(H(y_1,y_2,b_{\perp}) - H(x_1,x_2,b_{\perp})\Big) \nonumber\\
& + \int_{x_2}^\infty dy_2 \Big(\frac{x_2^2}{(y_2-x_2)y_2^2}H(y_1,y_2,b_{\perp})
-\frac{x_2}{(y_2-x_2)y_2}H(x_1,x_2,b_{\perp}) \Big) \Big] \nonumber\\
&+ \Big(C_F - \frac{C_A}{2} \Big) \int_{-\infty}^\infty dy_2 \Big[ -\theta(x_1-y_2)\frac{x_1(x_1^2+x_1(x_2-y_2)-2 x_2y_2)}{(x_1+x_2)^2 y_2^2}
\nonumber\\&
- \theta(x_1-y_2)\frac{x_2^2}{(x_1+x_2)^2 y_1}  \Big]
H(y_1,y_2,b_{\perp})  \Big\}
 .
\end{align}
And
\begin{align}
&\mathbb{K}_B H(x_1,x_2,b_{\perp})
\nonumber\\
=& \frac{ \alpha_s}{2\pi} \Big\{ \Big[\frac{3}{2}C_F + \Big( C_F-\frac{C_A}{2} \Big)\ln\frac{|x_1+x_2|}{|x_1|} + \frac{C_A}{2}\ln \frac{|x_1+x_2|}{|x_2|} +\Big(C_F-\frac{C_A}{2}\Big)i\pi \Big]H(x_1,x_2,b_{\perp})
\nonumber\\
& + \frac{C_A}{2} \Big[
\int^{x_1}_{-\infty} dy_1 \frac{x_1}{(x_1-y_1)y_1} \Big(H(y_1,y_2,b_{\perp}) - H(x_1,x_2,b_{\perp})\Big) \nonumber\\
& + \int_{x_2}^\infty dy_2 \Big(\frac{x_2^2}{(y_2-x_2)y_2^2}H(y_1,y_2,b_{\perp})
-\frac{x_2}{(y_2-x_2)y_2}H(x_1,x_2,b_{\perp}) \Big) \Big] \nonumber\\
&- \Big(C_F - \frac{C_A}{2} \Big) \int_{-\infty}^\infty dy_2  \theta(x_1-y_2)\frac{x_1}{ y_2^2}
H(y_1,y_2,b_{\perp})  \Big\}.
\end{align}

4. For $x_1<0, x_2>0, x_3<0$, we have
\begin{align}
&\mathbb{K}_A H(x_1,x_2,b_{\perp})
\nonumber\\
=& \frac{ \alpha_s}{2\pi} \Big\{ \Big[\frac{3}{2}C_F + \Big( C_F-\frac{C_A}{2} \Big)\ln\frac{|x_1+x_2|}{|x_1|} + \frac{C_A}{2}\ln \frac{|x_1+x_2|}{|x_2|} -\frac{C_A}{2}i\pi \Big]H(x_1,x_2,b_{\perp})
\nonumber\\
& + \frac{C_A}{2} \Big[\int_{-\infty}^\infty dy_2 \Big[ -\theta(y_2-x_2) \frac{x_1( x_1 + 2x_2)}{(x_1+x_2)^2 y_1}
 + \theta(y_2-x_2)  \frac{x_2^2(x_1+ x_2 + y_2)}{(x_1+x_2)^2 y_2^2}   \Big]H(y_1,y_2,b_{\perp})
  \nonumber\\
& +\int_{-\infty}^{x_1} dy_1 \frac{x_1}{(x_1-y_1)y_1} \Big(H(y_1,y_2,b_{\perp}) - H(x_1,x_2,b_{\perp})\Big) \nonumber\\
& + \int_{x_2}^\infty dy_2 \Big(\frac{x_2^2}{(y_2-x_2)y_2^2}H(y_1,y_2,b_{\perp})
-\frac{x_2}{(y_2-x_2)y_2}H(x_1,x_2,b_{\perp}) \Big) \Big] \nonumber\\
&+ \Big(C_F - \frac{C_A}{2} \Big) \int_{-\infty}^\infty dy_2 \Big[ -\theta(x_1-y_2)\frac{x_1(x_1^2+x_1(x_2-y_2)-2 x_2y_2)}{(x_1+x_2)^2 y_2^2}
\nonumber\\&
- \theta(x_1-y_2)\frac{x_2^2}{(x_1+x_2)^2 y_1}  \Big]
H(y_1,y_2,b_{\perp})  \Big\}
 .
\end{align}
And
\begin{align}
&\mathbb{K}_B H(x_1,x_2,b_{\perp})
\nonumber\\
=& \frac{ \alpha_s}{2\pi} \Big\{ \Big[\frac{3}{2}C_F + \Big( C_F-\frac{C_A}{2} \Big)\ln\frac{|x_1+x_2|}{|x_1|} + \frac{C_A}{2}\ln \frac{|x_1+x_2|}{|x_2|} -\frac{C_A}{2}i\pi \Big]H(x_1,x_2,b_{\perp})
\nonumber\\
& + \frac{C_A}{2} \Big[
\int^{x_1}_{-\infty} dy_1 \frac{x_1}{(x_1-y_1)y_1} \Big(H(y_1,y_2,b_{\perp}) - H(x_1,x_2,b_{\perp})\Big) \nonumber\\
& + \int_{x_2}^\infty dy_2 \Big(\frac{x_2^2}{(y_2-x_2)y_2^2}H(y_1,y_2,b_{\perp})
-\frac{x_2}{(y_2-x_2)y_2}H(x_1,x_2,b_{\perp}) \Big) \Big] \nonumber\\
&- \Big(C_F - \frac{C_A}{2} \Big) \int_{-\infty}^\infty dy_2  \theta(x_1-y_2)\frac{x_1}{ y_2^2}
H(y_1,y_2,b_{\perp})  \Big\}.
\end{align}

5. For $x_1<0, x_2<0, x_3<0$, we have
\begin{align}
&\mathbb{K}_A H(x_1,x_2,b_{\perp})
\nonumber\\
=& \frac{ \alpha_s}{2\pi} \Big\{ \Big[\frac{3}{2}C_F + \Big( C_F-\frac{C_A}{2} \Big)\ln\frac{|x_1+x_2|}{|x_1|} + \frac{C_A}{2}\ln \frac{|x_1+x_2|}{|x_2|} \Big]H(x_1,x_2,b_{\perp})
\nonumber\\
& + \frac{C_A}{2} \Big[\int_{-\infty}^\infty dy_2 \Big[ -\theta(y_2-x_2) \frac{x_1( x_1 + 2x_2)}{(x_1+x_2)^2 y_1}
 - \theta(x_2-y_2)  \frac{x_2^2(x_1+ x_2 + y_2)}{(x_1+x_2)^2 y_2^2}   \Big]H(y_1,y_2,b_{\perp})
  \nonumber\\
& +\int_{-\infty}^{x_1} dy_1 \frac{x_1}{(x_1-y_1)y_1} \Big(H(y_1,y_2,b_{\perp}) - H(x_1,x_2,b_{\perp})\Big) \nonumber\\
& + \int^{x_2}_{-\infty} dy_2 \Big(\frac{x_2^2}{(x_2-y_2)y_2^2}H(y_1,y_2,b_{\perp})
-\frac{x_2}{(x_2-y_2)y_2}H(x_1,x_2,b_{\perp}) \Big) \Big] \nonumber\\
&+ \Big(C_F - \frac{C_A}{2} \Big) \int_{-\infty}^\infty dy_2 \Big[ -\theta(x_1-y_2)\frac{x_1(x_1^2+x_1(x_2-y_2)-2 x_2y_2)}{(x_1+x_2)^2 y_2^2}
\nonumber\\&
+ \theta(y_2-x_1)\frac{x_2^2}{(x_1+x_2)^2 y_1}  \Big]
H(y_1,y_2,b_{\perp})  \Big\}
 .
\end{align}
And
\begin{align}
&\mathbb{K}_B H(x_1,x_2,b_{\perp})
\nonumber\\
=& \frac{ \alpha_s}{2\pi} \Big\{ \Big[\frac{3}{2}C_F + \Big( C_F-\frac{C_A}{2} \Big)\ln\frac{|x_1+x_2|}{|x_1|} + \frac{C_A}{2}\ln \frac{|x_1+x_2|}{|x_2|} \Big]H(x_1,x_2,b_{\perp})
\nonumber\\
& + \frac{C_A}{2} \Big[
\int^{x_1}_{-\infty} dy_1 \frac{x_1}{(x_1-y_1)y_1} \Big(H(y_1,y_2,b_{\perp}) - H(x_1,x_2,b_{\perp})\Big) \nonumber\\
& + \int^{x_2}_{-\infty} dy_2 \Big(\frac{x_2^2}{(x_2-y_2)y_2^2}H(y_1,y_2,b_{\perp})
-\frac{x_2}{(x_2-y_2)y_2}H(x_1,x_2,b_{\perp}) \Big) \Big] \nonumber\\
&- \Big(C_F - \frac{C_A}{2} \Big) \int_{-\infty}^\infty dy_2  \theta(x_1-y_2)\frac{x_1}{ y_2^2}
H(y_1,y_2,b_{\perp})  \Big\}.
\end{align}

Finally, the evolution equations with respect
to the scale $\zeta$ are the same for all domains, which read as
\begin{align}
\frac{d}{d\ln\zeta}H(x_1,x_2,b_{\perp})=& -\frac{\alpha_s C_F}{2\pi}\ln\frac{ b_T^2\mu^2}{4e^{-2\gamma_E}} H(x_1,x_2,b_{\perp}),
\end{align}
for $H\in \{ \tilde{f}^{\perp(1),\textrm{sub}}-i\tilde{g}^{\perp(1),\textrm{sub}}, \tilde{f}^{\perp(1),\textrm{sub}}+i\tilde{g}^{\perp(1),\textrm{sub}}, \tilde{h}^{\textrm{sub}}, \tilde{h}^{\perp(2),\textrm{sub}} \}$.

\providecommand{\href}[2]{#2}\begingroup\raggedright\endgroup

\end{document}